\documentstyle[aps,epsfig]{revtex}
\begin{document}
\title{Self force on charges in the spacetime of spherical shells}
\author{Lior M. Burko$^{a}$, Yuk Tung Liu$^{a}$, and Yoav Soen$^{b}$}
\address{$^{a}$Theoretical Astrophysics, California Institute of Technology,
Pasadena, California 91125
\\$^{b}$Department of Physics, Technion---Israel Institute of Technology,
32000 Haifa, Israel}
\date{\today}
\maketitle

\begin{abstract}
We study the self force acting on static electric or scalar charges inside
or outside a spherical, massive, thin shell. The regularization of the 
self force is done using the recently-proposed Mode Sum Regularization
Prescription. In all cases the self force acting on the charge is
repulsive. We find that in the scalar case the force is quadratic in
the mass of the shell, and is a second post-Newtonian effect. For the
electric case the force is linear in the shell's mass, and is a first
post-Newtonian effect. When the charge is outside the shell our results
correct the known zero self force in the scalar case or the known
repulsive, inverse-cubic force law in the electric case, for the finite
size of the shell. When the charge is near the center of the shell the
charge undergoes harmonic oscillations. 
\newline
\newline
PACS number(s): 04.25-g

\end{abstract}

\section{Introduction and Overview}\label{introduction}

The calculation of the self force acting on a particle in curved spacetime has become
recently exceedingly important, as the Laser Interferometer Space Antenna (LISA) is
currently planned
to fly as early as 2010. One of the most promising gravitational-wave sources for LISA is
extreme mass-ratio binaries, where the evolution in the last year of inspiral is strongly
affected by the self force acting on the inspiraling object. The long-term goal then, is
to compute the momentary self forces acting on a compact object, which is in a generic
orbit around a supermassive black hole. Once the self force is known, this force could be
included in the determination of the orbital evolution under radiation reaction.   

However, much is yet to be understood about the nature of self forces in curved spacetime 
even for problems which are significantly simpler
then the astrophysically-motivated ones. These problems, nevertheless, are
motivated by
being pedagogical on the one hand, and by illuminating some important points of principle,
which are relevant also for astrophysically-realistic problems, on the other hand. A
number of simple static configurations has thus been analyzed, including the self force
acting on scalar or electric charges held static in the spacetime of a Schwarzschild black
hole \cite{zelnikov-frolov-82,smith-will-80,wiseman-00,burko-cqg}, electric or magnetic
dipoles which are static outside a Schwarzschild black hole \cite{leaute-linet-84},
a static electric charge outside a Kerr black hole \cite{leaute-linet-82} or a Kerr-Newman
black hole \cite{lohiya-82}, a static electric charge in a spherically-symmetric
Brans-Dicke field \cite{linet-teyssandier-79}, and a static charge in the spacetime of a
cosmic string \cite{linet-86,smith-90}. In all these analyses the self force was found to
leading order in the charge of the particle, i.e., to order (charge)$^2$. In this Paper,
too, we calculate the self force to that order. 

Still, there are elementary configurations where
the self force acting on charges is as yet poorly understood. Specifically, little is
known on the self force acting on charges in the spacetime of spherical, massive  
(thin) shells.
This problem is interesting from a pedagogical point of view because the derivation of
the self force is simple, whether the charge is inside the shell (where spacetime is
locally flat) or outside the shell (where spacetime is locally Schwarzschild). However,
despite the simplicity, there are some non-trivial issues which are demonstrated
already in this simple context. First, when the charge is inside the shell, there is a
non-zero self force acting on it, even when the charge is static. Second, this non-zero
force exists although the charge moves along a geodesic of spacetime (a static worldline
is a geodesic in flat spacetime). Third, already in this simple context it is evident that
the Einstein Equivalence Principle is not satisfied by problems which involve self
interaction. Namely, two identical static particles, one in a globally-Minkowski spacetime
and one in a locally-Minkowski spacetime experience different forces: the former a zero
self force and the latter a non-zero self force, although the geometries of the local
neighborhoods of the two particles are exactly identical. [The self force is considered
here to order (charge)$^2$. Under this assumption, the geometry is a fixed background,
and is given by the solution of the Einstein equations in the absence of the charge.
When higher order terms, of order (charge)$^4$ or higher, are considered, the
geometries in the local neighborhoods of the particles would no longer be identical, 
because of the different ways in which the particles' fields couple to the geometry in
the two cases.] Because the self force couples to the
charge of the particle in a way which depends on the type of charge, the worldline of a
particle  which carries one type of charge deviates from the worldline of a particle which
carries a different type of charge. The reason for this failure of the Equivalence
Principle is obvious: the Equivalence Principle relates to the local
neighborhood of the particle. The self force, however, is affected 
by the boundary conditions on the surface of discontinuity (the shell), and these
far-field conditions affect the near-field of the particle such that a self force arises.
In problems where there is non-trivial dynamics, this effect comes about by the scattering
of the tails of the field off the spacetime curvature, and this scattering occurs also at
arbitrary large distances from the position of the particle in the remote past. In
this sense the self force is a non-local effect, 
which transcends the domain of applicability of
the Equivalence Principle.  
The Equivalence Principle is transcended also when the particle is outside the spherical
shell. Although spacetime outside a spherical shell of mass $M$ is identical to the
spacetime outside a black hole of the same mass $M$ (by virtue of Birkhoff's theorem), the
self forces acting on two identical particles, one in a globally-Schwarzschild spacetime
and the other in a locally-Schwarzschild spacetime, are different. In principle, one could
infer on (classes of) spherical mass distributions by solving the inverse problem, and
finding the spherical mass distribution which induces a measured self force. An important
point of principle then, is that the interior of the source for the gravitational field is
important for the orbital evolution of particles. As noted by Ryan \cite{ryan}, the
orbital evolution under radiation reaction would enable us to map the spacetime around a
black hole (or, possibly, some other dense objects). In addition, it would allow us also
to infer on the interior of the source, its equation of state, its density distribution,
et cetera. 

The problem of the self force in the spacetime of a spherical massive shell was first
considered by Unruh, who studied the case of a static electric
charge $q$ inside a spherical shell of mass $M$ and radius $R$ \cite{unruh-76}.
Unruh found that there was a radial
non-zero force acting
on the charge. [The related problem of a point charge, coupled to a massive (Proca)
vector field, within a spherical shell was studied by L\'eaut\'e and Linet
\cite{leaute-linet-85}.]  Below, we shall recapitulate Unruh's result, but derive
it using the newly-proposed Mode Sum Regularization Prescription (MSRP) 
\cite{ori-unpublished,barack-ori-00}.  We shall also provide
more insight into the meaning of the result. We find that the self force acting on the
charge is directed toward the center of the sphere (the self force acts to repel the
charge from the sphere). When the charge is very close to the center, this induces
harmonic oscillations, with angular frequency of $\omega^2=\frac{1}{3}(G/c^2)q^2M/(m
R^4)$, to leading order in $M/R$, where $m$ is the particle's mass. Here, $G$ is
Newton's gravitational constant, and $c$ is the speed of light.   

Then, we also study the self force acting on
an electric charge $q$ outside a spherical shell. For the case where the source of
the
gravitational field is a Schwarzschild black hole, the self force  was found by Smith and
Will \cite{smith-will-80}, who found that there was a repelling self force, which was
given by $f_{\hat r}(r)=(G/c^2)q^2M/r^3$. Here, $q$ is the electric charge of the
particle, $M$ is the black hole's mass, and $r$ is the usual radial Schwarzschild
coordinate.  
(This expression for the force is exact in a reference frame of a freely-falling observer
who is instantaneously at rest at the position of the charge.) 
We find that when the source of the
gravitational force is extended, the Smith-Will force is corrected by a finite-size term,
which, to leading order in $M/r$, is of the same order in $G/c^2$ as the Smith-Will term. 
Specifically, when the source is a thin spherical shell of mass $M$, to the leading order
in $M/r$, the self force is given by 
$f_{\hat r}=(G/c^2)q^2M/r^3[1+\frac{2}{3}(R/r)^2+O(R/r)^4]$, where $R$ is the radius of
the shell. For $r \gg R$ (when the charge is very far from the shell) this correction is
very small. However, when the charge is very close to the shell, higher-order terms in
$(R/r)^2$ become comparable with the Smith-Will term, such that the correction terms
may become more important than the leading-order term. We find that near the shell,
the self force becomes very large. (In fact, approaching the shell we find that the
self force diverges. We shall study that effect in detail, and show that when the charge
is very close to the shell our mathematical model breaks down, such that the physical
self force is always finite.)

Next, we calculate the self force acting on a particle endowed with scalar charge $q$. For
the case where the charge is inside the shell we again find a self force which directs
toward the center. However, this force is quadratic in $M/R$, and is of order $G^2/c^4$.
When the charge is near the center, the self force acting on it is again that of an 
harmonic oscillator, with angular frequency 
$\omega^2=\frac{1}{15}(G^2/c^4)q^2M^2/(m R^5)$, to the leading order in $M/R$. When the
scalar charge is outside the spherical shell, we find a correction to the well known
result of a zero self force which is the case when the source for the gravitational field
is a black hole \cite{zelnikov-frolov-82,wiseman-00,burko-cqg}. To leading
order in $M/r$, this self force is given by 
$f_{\hat r}(r)=\frac{1}{3}(G^2/c^4)q^2(M^2/r^5)R[1+O(R/r)^2]$. 
We find that when the
source of the gravitational field is extended, there
is a finite-size correction also for this case. However, because the zero-size force
happens to vanish, the finite-size effect is always larger than the zero-size effect
(unlike the electric charge case).   

The regularization prescription we use in the calculation of the self force is 
based on Ori's MSRP 
\cite{ori-unpublished,barack-ori-00}, which is an application of the axiomatic
Quinn-Wald approach \cite{quinn-wald-97} and the approach of 
Mino, Sasaki, and Tanaka \cite{mino-sasaki-tanaka-97}. 
We note, that although the regularization procedures
used in Refs.\ \cite{leaute-linet-84,leaute-linet-82,lohiya-82,linet-teyssandier-79,linet-86} 
yield the correct results for the regularized
self force, they are hard to generalize to cases were exact solutions are unknown, or where
there is non-trivial dynamics. MSRP does not share this difficulty, and the regularization
using MSRP is independent of the existence of an exact solution. 
Next, we describe MSRP very succinctly.
More
details are available in Refs. \cite{barack-ori-00,barack}. We note that MSRP has been
developed in detail only for scalar charges, but it is likely that the approach is
applicable in general also for higher-spin fields. 
The contribution to the physical self force from
the tail part of the Green's function can be decomposed into stationary Teukolsky modes,
and then summed over the frequencies $\omega$ and the azimuthal numbers $m$.
The self force equals then the limit $\epsilon\to 0^-$ of the
sum over all $l$ modes, of the difference between the force sourced by
the entire worldline (the bare force ${^{\rm bare}}F_{\mu}^{l}$) and
the force sourced by the half-infinite worldline 
to the future of $\epsilon$, where the particle has proper time
$\tau=0$, and $\tau=\epsilon$ is an event along the past ($\tau<0$) worldline. 
Next, we seek a regularization function $h^{l}_{\mu}$ which is independent of
$\epsilon$, such that the series $\sum_{l}({^{\rm bare}}F_{\mu}^{l}-h^{l}_{\mu})$
converges. Once such a function is found, the regularized self force is then given by
${^{\rm ren}}F_{\mu}=\sum_{l}({^{\rm bare}}F_{\mu}^{l}-h^{l}_{\mu})-
d_{\mu}$, where $d_{\mu}$ is a finite valued function. MSRP then shows,
from a local integration of the Green's function, that the regularization
function $h^{l}_{\mu}=a_{\mu}l+b_{\mu}+c_{\mu}l^{-1}$. For several cases, which
have already been studied,  
MSRP yields the values of the functions $a_{\mu}, b_{\mu}, c_{\mu}$
and $d_{\mu}$ analytically. Alternatively, $a_{\mu}, b_{\mu}$, and $c_{\mu}$ (but not
$d_{\mu}$) can also be found
from the large-$l$ behavior of ${^{\rm bare}}F_{\mu}^{l}$. 
In addition to the (non-local) contribution of the tail's part
of the Green's function to the self force, there are also two additional, local terms: an
Abraham-Lorentz-Dirac type term, and a term which couples to Ricci curvature and which
preserves conformal invariance. For simple
cases (including those we are interested in here) it was found that the finite term
$d_{\mu}$ equals the sum of the two local contributions \cite{barack,burko-amaldi}, such
that the total radiation-reaction force can be found directly from the large-$l$
behavior of the modes of the bare force. MSRP has already been applied successfully
for a number
of cases: static scalar or electric charges in the spacetime of a Schwarzschild black hole
\cite{burko-cqg}, a scalar charge in uniform circular orbit around a Schwarzschild black
hole \cite{burko-prl}, and a scalar charge plunging radially into a Schwarzschild black
hole \cite{barack-burko}. There is also strong evidence that MSRP is applicable also for
electric charges
\cite{burko-ajp,burko-cqg,burko-amaldi}. Recently, a closely-related regularization
procedure, based on Riemann $\zeta$-function regularization, was applied by Lousto for
the case of a point mass falling radially into a Schwarzschild black hole
\cite{lousto-00}. Lousto considered an orbit which is geodesic in the absence of 
radiation-reaction effects, and computed the first-order correction of the spacetime 
metric and orbit. It is presently unclear, however, how to extend Lousto's method for 
non-geodesic orbits. 

The organization of this Paper is as follows. 
In Section \ref{sec:scalar} we study the self force acting
on a scalar charge inside (\ref{sec:scalar-in}) and outside (\ref{sec:scalar-out}) a
spherical shell. In Section \ref{sec:electric} we consider the
self force on an electric charge, both inside (\ref{sec:electric-in})
and outside (\ref{sec:electric-out}) the shell, and in Section \ref{heuristic} make
some heuristic comments regarding the physical origin of the self force for the
problem in question. 

\section{Scalar charge}\label{sec:scalar}

Consider a thin uniform spherical shell of mass $M$ and radius $R$. In the spacetime
of this shell we place a static particle. In this section the particle is endowed
with a scalar charge $q$. (In Section \ref{sec:electric} we consider the case
where the particle is electrically charged.) We consider the scalar field to be a
linearized, test field  in the geometry of the shell's spacetime. That is, the field
is uncoupled to the geometry. This simplification allows us to obtain the self
force to order $q^2$. The particle is static at radius $r=r_0$, with either 
$r_0<R$ (in subsection \ref{sec:scalar-in}) or $r_0>R$ (in subsection
\ref{sec:scalar-out}) (see Fig.\ \ref{fig1}).
The fixed background geometry is described by the metric 
\begin{eqnarray}
  ds^2 = \left \{ \begin{array}{ll}
  -\left( 1-\frac{2M}{R} \right) dt^2 + 
dr^2 + r^2 d\theta^2 + r^2 \sin^2 \theta
d\varphi^2 & r<R \\
  -\left( 1-\frac{2M}{r} \right) dt^2 + \left(1-\frac{2M}{r}\right)^{-1} dr^2
+ r^2 d\theta^2 + r^2 \sin^2 \theta d\varphi^2 & r>R \end{array} \right. \ .
\label{metric}
\end{eqnarray}
Here the radial coordinate $r$ is defined such that the surface area of the
2-sphere $r$=constant, $t$=constant is $4\pi r^2$, and $t$ is the (proper) time of a
static observer at infinity.  
The spacetime (\ref{metric}) is Schwarzschild outside the shell and Minkowski inside the 
shell. Note that $g_{tt} \rightarrow -1$ as $r \rightarrow \infty$, but 
$g_{tt} \neq -1$ inside the shell, although spacetime is (locally) Minkowski.

The linearized field equation for a minimally coupled, massless scalar field 
$\Phi$ is given by 
\begin{equation}
  \nabla_{\mu} \nabla^{\mu}\Phi(x^{\alpha})= -4\pi \rho (x^{\alpha}) \ ,
\label{sc:eq}
\end{equation}
where $\nabla_{\mu}$ denotes covariant differentiation compatible with the metric
(\ref{metric}). 
The charge density $\rho$ is given by
\begin{equation}
  \rho(x^{\mu}) = q \int_{-\infty}^{\infty} \,d\tau \frac{\delta^4 [ x^{\mu}-
z^{\mu}(\tau)]}{\sqrt{-g}} \ .
\label{charge-density}
\end{equation}
Here $q$ is the total charge, $\tau$ is the proper time, $g$ is the metric 
determinant, and $z^{\mu}$ is the worldline of the charge. Without loss of 
generality, we place the charge on the $z$-axis at $r=r_0$ and $\theta=0$. 

\subsection{Scalar charge inside a spherical shell}\label{sec:scalar-in}
\subsubsection{Derivation of the bare force}
In this section we shall study the case where the scalar charge is inside the shell, i.e.,
$r_0<R$. 
Then both 
$\rho$ and 
$\Phi$ can be decomposed into a sum over the Legendre polynomials according to 
\begin{eqnarray}
    \rho(r,\theta) &=& \frac{q \, \delta(r-r_0)}{4\pi r_0^2} \sum_{l=0}^{\infty} 
(2l+1) P_l(\cos \theta) \label{sc:rho} \\
   \Phi(r,\theta) &=& \sum_{l=0}^{\infty} \phi^l(r) P_l(\cos \theta) \ .
\label{sc:phi}
\end{eqnarray}
Note that because $\,dt/\,d\tau=1/\sqrt{-g_{tt}}$, the factor $\sqrt{-g_{tt}}$ in the
metric determinant $g$ is canceled. Also note that the series in Eq.\ (\ref{sc:rho}) 
diverges. This is indeed what is expected, because the particle is construed as pointlike. 
Obviously, the charge density of a pointlike particle diverges on its worldline.  

Substitution of Eqs. (\ref{sc:rho}) and (\ref{sc:phi}) into Eq. (\ref{sc:eq}) yields  
\begin{equation}
 \left \{ \begin{array}{ll}
     \phi^l_{,rr}+\frac{2}{r}\phi^l_{,r}-\frac{l(l+1)}{r^2}\phi^l =-q\frac{2l+1}
{r_0^2} \delta(r-r_0)  & \ \ \ r<R\\
 & \\
     \left( 1-\frac{2M}{r} \right) \phi^l_{,rr}+\frac{2}{r^2}(r-M)\phi^l_{,r}
-\frac{l(l+1)}{r^2}\phi^l = 0 & \ \ \ r>R
\end{array} \right. \ ,
\label{sc:eq1}
\end{equation}
where commas denote partial derivatives.
The solution for the inhomogeneous equation for $r<R$ is given by a linear combination of
the solutions which solve the corresponding homogeneous equation, i.e., a linear
combination of $r^l$ and $r^{-l-1}$.  
Outside the shell, where $r>R$, the solution is a linear combination of the Legendre 
functions $P_l(\frac{r}{M}-1)$ and $Q_l(\frac{r}{M}-1)$ \cite{burko-cqg}. Note that  
$P_l(z)$ diverges and $Q_l(z)$ vanishes as $z \rightarrow \infty$ for all values of $l$. 
(Except for $l=0$, as $P_0(z)=1$ for all values of $z$. However, as we require that all 
the individual modes of the field $\phi^l$ fall off at infinity ($r\to\infty$), this 
behavior of $P_0(z)$ is enough to rule it out as the relevant solution for large values 
of $r$.) We next require that $\phi^l(r)$ be regular both at
infinity and at the origin. We thus write the solution for Eq. (\ref{sc:eq1}) as  
\begin{equation}
  \phi^l(r)=\left \{ \begin{array}{lll}
	 A_l \ Q_l\left(\frac{r}{M}-1\right) & \ \ \ r \geq R & \ \ \  {\rm (region \ III)}
\\
 & \\
	 C_l  \ r^l  + D_l \ r^{-(l+1)} & \ \ \ r_0 \leq r \leq R & \ \ \  {\rm (region  
\ II)} 
\\
 & \\
	 B_l \ r^l & \ \ \ r \leq r_0 & \ \ \  {\rm (region \ I)}\end{array} \right. 
\ \ ,
\label{sc:field}
\end{equation}
where $A_l$, $B_l$, $C_l$, and $D_l$ are constants to be determined by 
matching conditions. Specifically, we require 
that $\phi^l(r)$ be continuous everywhere 
(in particular, continuous also across $r=r_0$ and $r=R$).  

Integrating Eq. (\ref{sc:eq1}) over $r$ from $r=r_0-\epsilon$ to
$r=r_0+\epsilon$ and taking the limit 
$\epsilon \rightarrow 0^+$, and using the continuity
of $\phi^l(r)$ across $r_0$, we find
\begin{equation}
  \lim_{\epsilon \rightarrow 0^+}\left[ \phi^l_{,r}(r_0+\epsilon)-
\phi^l_{,r}(r_0-\epsilon)\right]=-\frac{(2l+1)\, q}{r_0^2} \ .
\label{bcro}
\end{equation}                                       
The integration of Eq. (\ref{sc:eq1}) over $r$ from $r=R-\epsilon$ to
$r=R+\epsilon$ is complicated by the discontinuity of $g_{rr}$ across $r=R$. This
unnecessary complication can be removed by introducing the new radial coordinate $u(r)$,
defined by  
\begin{equation}
  u \equiv \left \{ \begin{array}{ll}
	r & \ \ \ r \geq R \\
	& \\
	\sqrt{1-\frac{2M}{R}}(r-R)+R & \ \ \ r \leq R 
\end{array} \right. \ .
\label{eq:u}
\end{equation}
With this new radial coordinate, the metric (\ref{metric}) becomes 
\begin{equation}
  ds^2 = -K(u) dt^2 + \frac{du^2}{K(u)} + r(u)^2 (d\theta^2 + \sin^2 \theta d\varphi^2) \ . 
\end{equation}
Here, $r$ is considered to be a function of $u$ obtained by the inverse transform 
of (\ref{eq:u}), and 
\begin{equation}
  K(u) = \left \{ \begin{array}{ll} 
	   \left( 1-\frac{2M}{u}\right) & \ \ \ u \geq R \\
	    & \\
	   \left( 1-\frac{2M}{R}\right) & \ \ \ u \leq R 
	   \end{array} \right. 
\end{equation}
In this new gauge, $g_{uu}$ is continuous everywhere (and so are the other metric
coefficients) although the gradients of the metric functions $g_{tt,u}$, $g_{uu,u}$, and
$r_{,u}$ are still
discontinuous at $u=R$. In terms 
of $u$, Eq.\ (\ref{sc:eq1}) becomes 
\begin{equation}
  K \phi^l_{,uu}+\left(\frac{2K}{r}\frac{dr}{du}+K_{,u}\right)\phi^l_{,u}
-\frac{l(l+1)}{r^2}\phi^l=0 \ \ \ 
\mbox{for }r \neq r_0 \ .
\label{feq:u}
\end{equation}
Recall now that $\phi^l$ is continuous everywhere. 
However, $\phi^l_{,u}$ may at the most be step-function discontinuous at $u=R$ 
(the strongest form of discontinuity it may have is a step-function discontinuity; 
however, below we find that it is, in fact, continuous)   
such that $\phi^l_{,uu}$ may, at the most, behave like a $\delta$-function
(however, below we find that it is, in fact, proportional to a
step-function discontinuity) . In addition, also $K_{,u}$ and
$r_{,u}$ are step-function discontinuous, but do not involve $\delta$-functions. 
Consequently, when we integrate (\ref{feq:u}) over $u$ from $u=R-\epsilon$ to
$u=R+\epsilon$ and take
the limit $\epsilon \rightarrow 0^+$, only the contribution of the first term can be
non-vanishing (because that is the only term which may involve a $\delta$-function), and is 
given, after integration by part, by
\[
  \lim_{\epsilon \rightarrow 0^+} \left( \left. K \phi^l_{,u}\right|_{R-\epsilon}^
{R+\epsilon}-\int_{R-\epsilon}^{R+\epsilon} K_{,u} \phi^l_{,u}\, \,du \right)
=0 \ .
\] 
The second term vanishes since it does not contain a $\delta$-function, which implies that
$\phi^l_{,u}$ is continuous at 
$u=R$. In view of (\ref{eq:u}), this means that  
\begin{equation} 
  \sqrt{1-\frac{2M}{R}} \phi^l_{,r}(R^+) = \phi^l_{,r}(R^-) \ .
\label{bcR}
\end{equation}
Here, $R^{\pm}$ denotes $\lim_{\epsilon\to 0^{+}}(R\pm\epsilon)$. 

The constants $A_l$, $B_l$, $C_l$ and $D_l$ in (\ref{sc:field}) are determined 
by the matching conditions of $\phi^l$, which are (to reiterate): (1) $\phi^l$ 
is continuous everywhere; (2) $\phi^l$ vanishes at $r=\infty$ and is finite at 
$r=0$ [these two conditions have been taken into account in writing Eq.\ 
(\ref{sc:field})]; (3) $\phi^l_{,r}$ is discontinuous at $r=r_0$ and $r=R$ 
according to Eqs.\ (\ref{bcro}) and (\ref{bcR}). 
The results are 
\begin{mathletters} 
\begin{eqnarray}
   A_l &=& q \frac{(2l+1) {M\over R}}{l{M\over R}\, Q_l-\sqrt{1-{2M\over R}}\, Q'_l}
\frac{r_0^l}{R^{l+1}} \\
   C_l &=& -q\frac{r_0^l}{R^{2l+1}}E^s_l \\
   D_l &=& q r_0^l \\
   B_l &=& \frac{q}{r_0^{l+1}}+C_l \ \ ,
\end{eqnarray}   	
\end{mathletters}
where $Q'_l$ is the derivative of $Q_l$ with respect to its argument and the 
argument of both $Q_l$ and $Q'_l$ is $(R/M)-1$, and 
\begin{equation}
   E^s_l = \frac{\sqrt{1-{2M\over R}}\, Q'_l + (l+1){M\over R}\, Q_l}
{\sqrt{1-{2M\over R}}\, Q'_l - l{M\over R}\, Q_l} \ .
\end{equation}
Collecting our results, the field $\Phi$ is given by
\begin{equation}
\Phi(r,\theta) = \left \{ \begin{array}{lll}
        \sum_{l=0}^{\infty} A_l Q_l\left({r \over M}-1\right) P_l(\cos \theta) &
	 \ \ \ r \geq R & \ \ \ {\rm (region \ III)}  \\  
        & \\  
	\sum_{l=0}^{\infty} q {r_0^l \over r^{l+1}} P_l(\cos \theta) 
-\sum_{l=0}^{\infty} \frac{q}{R} ({r_0 \over R})^l
({r \over R})^l E^s_l P_l(\cos \theta) & \ \ \ r_0 \leq r \leq R 
& \ \ \ {\rm (region \ II)}  \\
        & \\
\sum_{l=0}^{\infty} q {r^l \over r_0^{l+1}} P_l(\cos \theta)  
-
\sum_{l=0}^{\infty} \frac{q}{R} ({r_0 \over R})^l
({r \over R})^l E^s_l P_l(\cos \theta)  & \ \ \ r \leq r_0 & \ \ \ {\rm (region \ I)}     
\end{array} \right. \ \ .
\label{sc:sc}
\end{equation}         

We are interested in the self force acting on the charge, which results from 
the field in the neighborhood of the particle, i.e., the field around $r=r_0$. As
$r_0<R$, the field in the neighborhood of the charge is given by 
\begin{equation}
\Phi(r<R,\theta) =
\sum_{l=0}^{\infty} q \left[ {r^l \over r_0^{l+1}}
\Theta(r_0-r)+{r_0^l \over r^{l+1}}\Theta(r-r_0)\right] P_l(\cos \theta)
-
\sum_{l=0}^{\infty} \frac{q}{R} \left({r_0 \over R}\right)^l
\left({r \over R}\right)^l E^s_l P_l(\cos \theta) \ \ .
\end{equation} 
Here, $\Theta(x)$ is the Heaviside step function, i.e., $\Theta(x)=1$ if $x>0$ and
$\Theta(x)=0$ if $x<0$.  
To find the self force, we next calculate the force according to 
$f_{\mu}=q\nabla_{\mu}\Phi$, and evaluate this expression at the position of the particle,
i.e., at $r=r_0$. (We note that the alternative force law which is frequently used, i.e., 
$f_{\mu}=q(\nabla_{\mu}\Phi+u_{\mu}u^{\nu}\nabla_{\nu}\Phi)$, is likely not to be derivable
from an action principle without introduction of non-linear coupling terms
\cite{wiseman-notes}. The simpler force law we use was recently justified by Quinn
\cite{quinn} using stress-energy considerations. Notice, however, that for the static
particle we assume here the two force laws coincide.) 
Clearly, only the radial component of the force $f_{\mu}$ is non-zero.
The (bare) radial force is then given by 
\begin{equation}
f_{r}^{\rm bare}=-\sum_{l=0}^{\infty}\left[ 
\frac{q^2}{2r_0^2} + 
\left({q \over R}\right)^2  l\, E_l^s\,
\left({r_0 \over R}\right)^{2l-1}\right] \ . 
\end{equation}
Clearly, $f_{r}^{\rm bare}$ diverges. 
\subsubsection{Regularization of the bare force}
In order to find the regularized, physical self force
we next use MSRP. First, we find the regularization function $h_r=a_rl+b_r+c_rl^{-1}$. Note
that
the functions $a_r,b_r$, and $c_r$ are found from a local analysis of the Green's function,
in
the neighborhood of the worldline. However in the neighborhood of the worldline
spacetime
is Minkowski. For a static scalar particle in Minkowski spacetime the values of 
$a_r,b_r$, and $c_r$ are known, and are given by $a_r=0=c_r$, and $b_r=-q^2/(2r_0^2)$. 
(Note that these values were obtained for the gauge where $g_{tt}=-1$. 
However, re-definition of the coordinate $t$ does not change the values of these
functions. Another point to be made here is that we use here the ``averaged'' 
value for $a_r$, which vanishes. The analysis can also be carried out using the
``one-sided'' values $a_r^{\pm}$, or any of their linear combinations. However,
MSRP guarantees the same result for the regularized self force whatever the choice
for $a_r$ may be. For more details see Ref.\ \cite{barack-ori-00}.) 
Next, for a static
scalar particle in Minkowski $d_r=0$, such that the regularized radial self force 
is given by 
\begin{eqnarray}
f_r&=&-\sum_{l=0}^{\infty}\left[
\frac{q^2}{2r_0^2} +
\left({q \over R}\right)^2  l\, E_l^s\,
\left({r_0 \over R}\right)^{2l-1}+b_r \right] \nonumber \\
&=& -\left({q \over R}\right)^2 \sum_{l=0}^{\infty} l\, E_l^s\,
\left({r_0 \over R}\right)^{2l-1} \ \ .
\label{sc:sf}
\end{eqnarray}   
This expression for $f_r$ is guaranteed by MSRP to be finite, and also to be the correct,
physical, total self force. 
\subsubsection{Properties of the regularized force}
It can be shown that $E^s_l$ is always positive, such that the self 
force directs towards the origin. In terms of the physical
forces present in the problem, this force arises from the sphere, and acts to repel the
charge from it. To gain
more insight into this result, we next expand $E_l^s$ in $M/R$. This expansion is given by 
\begin{equation}
  E_l^s=\frac{l+1}{2(2l+1)(2l+3)}\left({M\over R}\right)^2 +
O\left({M\over R}\right)^3 \ ,
\end{equation}
such that the self force becomes
\begin{equation}
   f_r = -\left({q\over R}\right)^2 \left({M\over R}\right)^2 \left[ 
\sum_{l=1}^{\infty} \frac{l(l+1)}{2(2l+1)(2l+3)}\left({r_0\over R}\right)^{2l-1} +
O\left({M\over R}\right) \right] \ .
\end{equation}
To the leading order in $M/R$ the self force is given by 
\begin{equation}
   f_r = -q^2 \frac{G^2}{c^4}
\frac{M^2}{R^4}
\left[
\frac{3-x_0^2}{16x_0^3 (1-x_0^2)}-\frac{3+x_0^2}{32x_0^4}\ln \left({1+x_0 \over
1-x_0}\right) \right]   \ ,
\label{sc:lin}
\end{equation}
where $x_0\equiv r_0/R$. In Eq. (\ref{sc:lin}) we re-introduced Newton's gravitational
constant $G$, and the speed of light $c$. Equation (\ref{sc:lin}) implies that this force 
is a 2nd post-Newtonian effect. 

In this small $M/R$ limit, we further consider two cases. The first case is the 
charge being close to the center. The self force is 
\begin{equation}
  f_r=-{1\over 15}q^2\frac{G^2}{c^4}\frac{M^2}{R^5}r_0 \left[1+ 
O\left(r_0\over R \right)^2 \right] \ .
\label{sc:force-law}
\end{equation}
The charge then oscillates under this self force at angular frequency 
\begin{equation}
  \omega^2=\frac{1}{15}q^2\frac{G^2}{c^4}\frac{M^2}{m R^5} \ ,
\end{equation}
where $m$ is the particle's mass. We note that although the force law 
(\ref{sc:force-law}) corresponds to the case where the particle is static, when the
deviation from the center is small enough, the velocity during the oscillations is
small enough such that the correction to the static force-law is negligible, and the
oscillations' frequency is, to first order, unchanged.

The next case we consider is the case where the charge approaches the shell, i.e. 
$x_0 \rightarrow 1^-$. To the leading orders in $1-x_0$  the self force becomes
\begin{equation}
  f_r=-q^2 \frac{G^2}{c^4} \frac{M^2}{R^4} \left \{ \frac{1}{16(1-x_0)}+
{1\over 8}\ln(1-x_0)+{9\over 32}-{\ln 2\over 8} +O[ (1-x_0)\ln(1-x_0)]\right \} \ ,
\label{sc:lim-in}
\end{equation}
which implies that the self force increases when the particle is closer to the
shell. In fact, in the limit $x_0 \rightarrow 1^-$ this self force diverges. Recall
that this self force is supposedly the regularized, physical self force, which
should be everywhere finite, even at the surface of discontinuity. Below, in Section
\ref{heuristic} we discuss the origin for the self force. We show there that this
origin is the gravitational interaction between the charge's field and the shell's
mass elements. (In fact, in Section \ref{heuristic} the analysis is done to only
linearized order in $M$, such that in the scalar case, where the force is quadratic in
$M$, it yields a zero effect; however, for the electric case, where the self force
is linear in $M$, it yields a result in total agreement with our analysis below in
Section \ref{sec:electric}.) Due to the spherical symmetry of the shell, clearly
only the parts of the field which are external to the shell contribute to the total
interaction. The closer the particle to the shell, the stronger the field in the
exterior of the shell in the neighborhood of the latter, such that indeed one
expects the interaction to be stronger. The divergence we find in the coincidence
limit is a shortcoming of the mathematical model we use. Recall that we treat both
the shell and the particle as mathematical $\delta$-functions. In actuality, one
should endow both with finite extensions, which would remove this divergence. The 
$\delta$-functions model fails because it is incompatible with the assumption of
staticity which we make. Specifically, the pointlike particle has some finite energy
density and mass. In the electric analog, the electrically-charged particle has an
electrostatic field which gravitates, and a mass. 
For an electron the charge is much larger
than the mass, such that the latter can be ignored for the 
purpose of this discussion. The
gravitational effect of the electrostatic 
field acts then to repel the shell. Approaching the
shell this repulsion grows unboundedly, such that at some finite distance the internal
stresses in the shell will no longer hold, and the geometry becomes dynamic. Hence, the
assumption of staticity fails in the 
coincidence limit. When the particle is endowed with a
large mass (but still considered as pointlike), it is, in fact, 
a tiny black hole. When that black
hole is put too closely to the shell, the shell's internal stresses would no longer
be able to hold against the pull of the particle's gravity, and collapse. In our
mathematical analysis we assume that the entire configuration is static, and this
assumption fails in the coincidence limit. An analogous discussion can be made also
for a
particle endowed with a scalar charge. 

Next we consider the case when $M/R$ is not small. Fig.\ \ref{fig:scin} shows the 
magnitude of the self force in unit of $q^2G^2/(c^4 R^2)$ as a function 
of $R/M$ calculated by the full expression (\ref{sc:sf}) (solid line) and the small $M/R$ 
expression (\ref{sc:lin}) (dashed line). (This full expression, and similarly the full
expressions computed below, is computed using the numerical method outlined in Ref.\ 
\cite{burko-cqg}.) 
The charge is placed at $r_0/R=1/2$. 
It can be seen that the self force increases as $R/M$ decreases and 
the linear expression (\ref{sc:lin}) deviates from the full expression. It can be 
shown that the self force approaches 
\begin{equation}
  f_r \rightarrow -\frac{q^2}{R^2} {G^2\over c^4} \frac{x_0}{(1-x_0^2)^2}
\label{sc:asyin}
\end{equation}
when $R/M\rightarrow 2$.
\subsubsection{Alternative regularization procedure}
We remark that the regularization procedure, which we performed using MSRP, can be
justified by a more straightforward, albeit less robust approach. We can make direct use of
the scalar-field Comparison Axiom of Quinn \cite{quinn} (which is modeled after the
Comparison Axiom for electric and gravitational fields of Quinn and Wald, for which
plausibility arguments were given \cite{quinn-wald-97}). The scalar-field
Comparison Axiom states the following (see \cite{quinn} for more details):
{\it Consider two points, $P$ and
$\tilde{P}$, each lying on time-like worldlines in possibly different spacetimes which
contain Klein-Gordon fields $\phi$ and $\tilde{\phi}$ sourced by particles of charge $q$ on
the worldlines. If the four-accelerations of the worldlines at $P$ and $\tilde{P}$ have the
same magnitude, and if we identify the neighborhoods of $P$ and ${\tilde P}$ via the
exponential map such that the four-velocities and four-accelerations are identified 
via Riemann normal coordinates, then the difference between the scalar forces
$f_{\mu}$ and ${\tilde f}_{\mu}$ is given by the
limit $x\to 0$ of the field gradients, averaged over a sphere at geodesic distance $x$
from the worldline at $P$, i.e.,} 
\begin{equation}
f_{\mu}-{\tilde f}_{\mu}=\lim_{x\to 0} q\left<\nabla_{\mu}\phi-{\tilde
\nabla}_{\mu}{\tilde
\phi}\right>_x \ .
\label{sc:ax}
\end{equation}
We note that this axiom assumes a nearly trivial form for the case in question: the local
neighborhood of the particle in question and of a similar particle in a
(globally-)Minkowski spacetime are identical. (It is only the far-away properties of
spacetime which are different for the two spacetimes.) Another remark is that we do not
need to average here over directions, as the forces in our case is direction independent. 
Setting our minds to use this
Comparison Axiom, we write the scalar field (\ref{sc:sc}) as 
\begin{equation}
  \Phi(r,\theta) = \left \{ \begin{array}{ll}
	\sum_{l=0}^{\infty} A_l Q_l\left({r \over M}-1\right)
	P_l(\cos \theta) & \ \ \ r \geq R \\
	& \\
	\Phi_c(r,\theta) + \Delta \Phi_s(r,\theta) & \ \ \ r \leq R 
\end{array} \right. \ \ ,
\end{equation}
where
\begin{eqnarray}
  \Phi_c(r,\theta) &=& \sum_{l=0}^{\infty} q \left[ {r^l \over r_0^{l+1}}
\Theta(r_0-r)+{r_0^l \over r^{l+1}}\Theta(r-r_0)\right] P_l(\cos \theta) \cr
 &=& \sum_{l=0}^{\infty} q \frac{r_<^l}{r_>^{l+1}} P_l(\cos \theta) \cr
&=& \frac{q}{|\mbox{\boldmath $r$}-r_0 \mbox{\boldmath $\hat{z}$}|}
\label{coulumb} 
\end{eqnarray} 
is just the usual scalar field for a particle at rest in a globally-Minkowski spacetime,
and $\Delta \Phi_s(r,\theta)$ is a term which corrects for the finite 
size of the Minkowski patch of spacetime, which is given by 
\begin{eqnarray}
  \Delta \Phi_s(r,\theta) = -\sum_{l=0}^{\infty} \frac{q}{R} 
\left({r_0 \over R}\right)^l 
\left({r \over R}\right)^l E^s_l P_l(\cos \theta) \ \ ,
\end{eqnarray}
Here $r_<=\min(r,r_0)$ and $r_>=\max(r,r_0)$. 
Hence inside the shell, the scalar field contains the Coulomb field $\Phi_c$ and 
a correction $\Delta \Phi_s$. We next identify $\Phi_c$ with ${\tilde \phi}$. The self
force
therefore arises only from $\Delta \Phi_s$, and is given by
\begin{eqnarray}
   f_r &=& q(\Delta \Phi_{s})_{,r} \\
   &=& -\left({q \over R}\right)^2 \sum_{l=1}^{\infty} l\, E_l^s\,
\left({r_0 \over R}\right)^{2l-1} \ ,
\end{eqnarray}
a result which is identical to Eq. (\ref{sc:sf}).                           
(Alternatively, we can argue that 
the self force on a static particle in a globally-Minkowski
spacetime is zero, such that $\Phi_c$ does not contribute to the self force. The self force
therefore arises only from $\Delta \Phi_s$.)

\subsection{Scalar charge outside the shell}\label{sec:scalar-out}
\subsubsection{Derivation of the bare force}
Next, we study the case where the scalar 
charge is outside the shell, i.e., the case where 
$r_0>R$. Now, the charge density (\ref{charge-density}) is decomposed as
\begin{equation}
  \rho (r,\theta) = \frac{q}{4\pi}\, \frac{\delta (r-r_0)}{r_0^2} \sqrt{1-{2M\over
r_0}}
\sum_{l=0}^{\infty} (2l+1)\, P_l(\cos \theta) \  \ ,
\end{equation}
[instead of (\ref{sc:rho})]
and the potential is decomposed as in (\ref{sc:phi}).
Then the scalar-field equation $\nabla_{\mu} \nabla^{\mu} \Phi=-4\pi \rho$
becomes
\begin{equation}
\left \{  \begin{array}{ll}
  \left( 1-{2M\over r}\right) \phi^l_{,rr}+{2\over r^2} (r-M) \phi^l_{,r}
-\frac{l(l+1)}{r^2} \phi_l = -q\frac{2l+1}{r_0^2}\delta(r-r_0)
\sqrt{1-{2M\over r_0}} & \ \ \ \ \ r>R \\  & \\
\phi^l_{,rr}+{2\over r} \phi^l_{,r}-\frac{l(l+1)}{r^2}\phi_l=0 & \ \ \ \ \
r<R
\end{array} \right. \ \ .
\label{sc-out:eq}
\end{equation}
Our solution here follows closely the solution of the preceding section. 

The $\delta$-function in Eq.\ (\ref{sc-out:eq}) gives one of the matching
conditions:
\begin{equation}
  \lim_{\epsilon \rightarrow 0^+} \left[ \phi^l_{,r}(r_0+\epsilon)-
\phi^l_{,r}(r_0-\epsilon) \right] = -q \frac{2l+1}{r_0^2}
\left( 1-{2M\over r_0}\right)^{-1/2} \ \ \ ,
\end{equation}
while the other matching conditions are the same as in the case when
the scalar charge is inside the shell. The solution is expressed as a
linear combination of Legendre functions $P_l$ and $Q_l$ with arguments 
$r/M-1$ for $r>R$, and is proportional to $r^l$ for $r<R$. Hence, 
\begin{equation}
  \phi^l(r) = \left \{ \begin{array}{lll}
        A_l Q_l \left({r\over M}-1\right) & \ \ \ \ \ r\geq r_0 
& \ \ \ \ {\rm (region \ III)} \\ & \\
        C_l P_l \left({r\over M}-1\right) + D_l Q_l \left({r\over
M}-1\right)
 & \ \ \ \ \ R \leq r \leq r_0 & \ \ \ \ {\rm (region \ II)} 
\\ & \\
        B_l r^l & \ \ \ \ \ r\leq R  & \ \ \ \ {\rm (region \ I)} 
\end{array} \right. \ \ ,
\end{equation}
where the constants $A_l$, $B_l$, $C_l$ and $D_l$ are determined by the
matching conditions. The result for the scalar field $\Phi(r,\theta)$ is then given by
\begin{equation}
\Phi(r,\theta) = \left \{ \begin{array}{lll}
        {q\over M} \sqrt{1-{2M\over r_0}}\sum_{l=0}^{\infty}
(2l+1) P_l(\cos \theta)\left[ P_l(z_0)+\tilde{E}^s_l Q_l(z_0) \right] Q_l(z)
& \ \ \ r \geq r_0 & \ \ \ {\rm (region \ III)}  \\ & \\
{q\over M} \sqrt{1-{2M\over r_0}}\sum_{l=0}^{\infty}    
(2l+1) P_l(\cos \theta)Q_l(z_0) 
\left[ P_l(z)+\tilde{E}^s_l Q_l(z)\right]          
& \ \ \ R \leq r \leq r_0 & \ \ \ {\rm (region \ II)}  \\
        & \\
q\frac{(2l+1)\, M}{R^{l+2}}
\sqrt{\frac{R(r_0-2M)}{r_0(R-2M)}}
\frac{Q_l(z_0)}
{ l{M\over R} Q_l(Z) - \sqrt{1-{2M\over R}} Q_l'(Z)} \ 
r^l \ P_l(\cos \theta)
  & \ \ \ r \leq R & \ \ \ {\rm (region \ I)}
\end{array} \right. \ \ .
\label{sc:sc1}
\end{equation}                      
Here, $z\equiv r/M-1$, $z_0\equiv r_0/M-1$, $Z\equiv R/M-1$, and we have used the 
Wronskian $W[P_n(z),Q_n(z)]=-1/(z^2-1)$ 
to simplify the expression. We are interested in the
field in the neighborhood of the particle. As $r_0>R$, the field in this neighborhood can
be written as 
\begin{eqnarray}
   \Phi(r>R,\theta) =
{q\over M} \sqrt{1-{2M\over r_0}}\sum_{l=0}^{\infty}
(2l+1) P_l(\cos \theta)\left[ P_l(z_0)Q_l(z)\Theta(r-r_0)+P_l(z)Q_l(z_0)\Theta(r_0-r)
+\tilde{E}^s_lQ_l(z_0)Q_l(z)\right], 
\end{eqnarray}
where $\tilde{E}^s_l$ is given by
\begin{equation}
   \tilde{E}^s_l = \frac{l{M\over R}\, P_l(Z)-\sqrt{1-{2M\over R}}\, P_l'(Z)}
{\sqrt{1-{2M\over R}}\,
Q_l'(Z)-{lM\over R}\, Q_l(Z)} \ ,
\label{sc:elout}
\end{equation} 
which can be shown to be positive for all $l$ and $M/R<1/2$.

To find the self force, we next calculate the force according to
$f_{\mu}=q\nabla_{\mu}\Phi$, and evaluate this expression at the position of the particle,
i.e., at $r=r_0$. Clearly, only the radial component of the force $f_{\mu}$ is non-zero.
The (bare) radial force is then given by
\begin{equation}
f_{r}^{\rm bare}=\frac{q^2}{M^2}\sqrt{1-\frac{2M}{r_0}}
\sum_{l=0}^{\infty}(2l+1) \left\{ 
\frac{1}{2}\left[P_l(z_0)Q_l'(z_0)+P_l'(z_0)Q_l(z_0)\right]+\tilde{E}^s_lQ_l(z_0)Q_l'(z_0)
\right\} \ .
\label{sc:bare}
\end{equation}                         
The series, when summed naively, diverges. 
\subsubsection{Regularization of the bare force}
To regularize the bare force 
 and extract the finite, physical
self force, we use MSRP. Again, only the radial self force is not trivial. 
The regularization function $h_r=a_rl+b_r+c_rl^{-1}$ can be found from a local analysis
near the particle, where spacetime is locally Schwarzschild. 
In Schwarzschild spacetime the
values of $a_r, b_r, c_r$, and $d_r$ are completely 
known for a static scalar particle, and
are given by $a_r=0=c_r$ (we again use here the ``averaged'' value for $a_r$), and
$b_r=-[q^2/(2r_0^2)][(1-M/r_0)/(1-2M/r_0)]$. 
In addition, it
is also known that $d_r=0$. In Ref. \cite{burko-cqg} it was shown that when MSRP is applied to
the  term proportional to the square brackets in Eq. (\ref{sc:bare}), the sum over all
modes vanishes. That is, when $f_{r}^{\rm bare}$ is written as a sum over modes, i.e., 
$f_{r}^{\rm bare}\equiv \sum_{l=0}^{\infty}f_{r}^{{\rm bare} \,\, l}$, the regularized
self force can be written as 
$f_{r}=\sum_{l=0}^{\infty}\left(f_{r}^{{\rm bare} \,\, l}-b_r\right)$. Because this series
converges, it can be split into two sums, the first of which vanishes \cite{burko-cqg}, such
that the regularized, physical self force in an orthonormal basis is given by 
\begin{equation}
   f_{\hat{r}}=\left( {q\over M}\right)^2 \left(1-{2M\over
r_0}\right)\sum_{l=0}^{\infty}
(2l+1) \tilde{E}^s_l Q_l \left({r_0\over M}-1\right) Q_l' \left({r_0\over
M}-1\right) \ \ .
\label{sc:sf2}
\end{equation}                
This is a repulsive force, as it always directs away from 
the center of the coordinates (and also away from the shell). This result
(\ref{sc:sf2}) corrects the zero self-force result (for the case where the source of
the gravitational field is a black hole
\cite{zelnikov-frolov-82,wiseman-00,burko-cqg}) for the finite size of the shell.   
\subsubsection{Properties of the regularized force} 
For small $M/R$, the leading order of Eq.\ (\ref{sc:sf2}) is given by 
to
\begin{equation}
   f_{\hat{r}} = \left( {q\over r_0}\right)^2 \left( {M\over r_0}\right)^2
\sum_{l=1}^{\infty} \frac{l(l+1)}{2(2l-1)(2l+1)}\left( {R\over
r_0}\right)^{2l-1}
\end{equation}
which can be written as 
\begin{equation}
   f_{\hat{r}} = q^2 \frac{G^2}{c^4} \frac{M^2}{r_0^4}
\left[ \frac{1+3y_0^2}{32y_0^2}\ln \left( {1+y_0\over
1-y_0}
\right)-\frac{1-3y_0^2}{16y_0(1-y_0^2)} \right] \ \ ,
\label{scout:lin}
\end{equation}
where $y_0\equiv R/r_0$. Here, we re-introduced Newton's constant $G$ and the speed of 
light $c$. As the charge approaches the shell ($y_0 \rightarrow 1^-$), 
\begin{equation}
  f_{\hat{r}}=q^2 \frac{G^2}{c^4} \frac{M^2}{r_0^4} \left \{ \frac{1}{16(1-y_0)}
-{1\over 8}\ln(1-y_0)-{3\over 32}+{\ln 2\over 8}+O[(1-y_0)\ln(1-y_0)]\right \} \ .
\label{sc:lim-out}
\end{equation}
The divergence which we find in the coincidence limit is again due to the simplified
model of the pointlike charge (see discussion above).  Note, that the result  
for the self force in the limit where the particle approaches the shell from the
outside (\ref{sc:lim-out}) is similar to the result
for the self force in the limit where the particle approaches the shell from the
inside (\ref{sc:lim-in}). In fact, the leading order terms (in the inverse of the
distance from the shell) of the two cases are identical (they only have opposite
signs, because the self force in both cases repels the particle from the shell).
However, the next terms (which are proportional to the logarithm of the distance
from the shell) are no longer identical (the relative signs are different) because
of the different direction of the shell's curvature with regard to the particle's
position.   

Fig.\ \ref{fig:scout} shows the self force $f_{\hat r}$ as a 
function of $R/M$ computed by the full expression (\ref{sc:sf2}) (solid line) and the 
small $M/R$ formula (\ref{scout:lin}). The charge is placed at $r_0/R=2$. The 
self force increases as $R/M$ decreases, reaches a maximum at $R/M \approx 2.3$ and 
drops to zero when $R/M \rightarrow 2$. It is easy to show analytically
from (\ref{sc:elout}) and (\ref{sc:sf2}) that the self force 
vanishes in the limit 
$R/M \rightarrow 2$, which coincides with the case when 
the shell is replaced by a Schwarzschild black hole. Of course, in that limit
the shell can no longer be static, and must implode instead. The spacetime
then is that of a Schwarzschild black hole with appropriate boundary
conditions at the event horizon, which are unaffected by the implosion of the
shell inside the black hole. That is the reason why the static shell gives
the right result (of a zero self force) in the limit of coincidence of its
radius with the Schwarzschild radius. 

Finally, Fig.\ \ref{fig:x0}a displays the magnitude of the self force $|f_{\hat r}|$  
as a function of $r_0/R$ for $R=2.5M$. We can see that 
the small $M/R$ expressions (dashed line) are no longer accurate in this 
case. The self force decreases as the charge moves away from the shell, but  
diverges at $r_0=R$ (see the above discussion).
\subsubsection{Alternative regularization procedure}
Again, the regularization prescription can be performed also by applying directly Quinn's
Comparison Axiom. This time, we choose the spacetime of a globally-Schwarzschild spacetime
(a Schwarzschild black hole) as the ``tilde'' spacetime. 
That is, we write the total scalar field as 
\begin{equation}
   \Phi(r,\theta) = \left \{ \begin{array}{ll}
  \Phi^{\rm sch}(r,\theta)+\Delta \Phi(r,\theta) & \ \ \ \ \ r \geq R \\ & \\
  B_l \ r^l \ P_l(\cos \theta) & \ \ \ \ \ r \leq R \end{array} \right. \ \ ,
\end{equation}
where
\begin{equation}
   B_l=q\frac{(2l+1)M}{R^{l+2}}\left(1-{2M\over r_0}\right)^{1/2}
\left(1-{2M\over R}\right)^{-1/2} \frac{Q_l \left({r_0\over M}-1\right)}
{ l{M\over R} Q_l \left( {R\over M}-1\right) - \sqrt{1-{2M\over R}} Q_l'
\left( {R\over M}-1\right)} \ \ ,
\end{equation}
and 
\begin{eqnarray}
  \Phi^{\rm sch}(r,\theta) &=& {q\over M} \sqrt{1-{2M\over
r_0}}\sum_{l=0}^{\infty}
(2l+1) P_l(\cos \theta) \left[ P_l \left({r_0\over M}-1\right)
Q_l \left({r\over M}-1\right) \Theta(r-r_0)\right. \nonumber \\ 
&+&\left.  P_l \left({r\over M}-1\right)
Q_l \left({r_0\over M}-1\right) \Theta(r_0-r) \right]
\end{eqnarray}
is the scalar field if the shell is replaced by a Schwarzschild black hole
of the same mass \cite{burko-cqg}, which we choose to be ${\tilde \phi}$, 
and
\begin{equation}
  \Delta \Phi(r,\theta)={q\over M} \sqrt{1-{2M\over
r_0}}\sum_{l=0}^{\infty}
(2l+1) \tilde{E}^s_l Q_l \left({r_0\over M}-1\right) Q_l \left({r\over M}-1\right)
P_l(\cos \theta) \ \ ,
\end{equation}
is the correction due to the finiteness of the shell's radius. Applying the Comparison
Axiom, the self force is given by 
\begin{eqnarray}
   f_r &=& q(\Delta \Phi)_{,r} \cr
   f_{\hat{r}} &=& \left( {q\over M}\right)^2 \left(1-{2M\over
r_0}\right)\sum_{l=0}^{\infty}
(2l+1) \tilde{E}^s_l Q_l \left({r_0\over M}-1\right) Q_l' \left({r_0\over
M}-1\right) \ \ , 
\end{eqnarray}
which is identical to Eq. (\ref{sc:sf2}). [This can also be stated, alternatively, in the
following way: 
Because the self force is zero when the source of the gravitational field is point-like,
the self force is given only by $f_r = q(\Delta \Phi)_{,r}$.]

\section{Electric Charge}\label{sec:electric}

The second type of charge we study is an electric charge $q$, which again, following the
preceding section, is at rest inside or outside a spherical shell of mass $M$. 
In this case, we look for the solution of the Maxwell equation 
\begin{equation}
  \nabla_{\nu} F^{\mu \nu} = 4\pi j^{\mu} \ \ ,
\label{el:eq}
\end{equation}
where $F_{\mu \nu}$ is the Maxwell field-strength tensor, which is antisymmetric 
and related to the 4-vector potential $A_{\mu}$ by $F_{\mu \nu} = A_{\nu,\mu}-
A_{\mu,\nu}$. Here, $j^{\mu}=\rho u^{\mu}$ is the four-current density, $\rho$ is
the  charge density, and $u^{\mu}$ is the 4-velocity of the charge. Since the charge
is static, the spatial components of $j^{\mu}$ vanish. From the spatial
components 
of Eq.\ (\ref{el:eq}), one can easily show that it is possible to choose a gauge 
so that all spatial components of $A_{\mu}$ vanish. 
The equation for $A_t$ is given by the temporal component of (\ref{el:eq}): 
\begin{equation}
  \frac{1}{\sqrt{-g}} (\sqrt{-g}\, g^{\nu \alpha}\, g^{tt}\, A_{t,\alpha})_{,\nu}
= -4\pi j^t \ ,
\label{el:eq2}
\end{equation}
with $j^t$ given by 
\begin{equation}
  j^t=q \int_{-\infty}^{\infty} u^t\, \frac{\delta^4 [x^{\mu}-z^{\mu}(\tau)]}
{\sqrt{-g}}\, d\tau \ .
\end{equation}

\subsection{Electric charge inside the shell}\label{sec:electric-in}
\subsubsection{Derivation of the bare force}
For the case where the electric charge is inside the shell, $r_0<R$. This case was studied
by Unruh in Ref.\ \cite{unruh-76}. 
The decomposition of 
$A_t$ and $j^t$ into modes yields 
\begin{eqnarray}
    A_t(r,\theta) &=& \sum_{l=0}^{\infty} \phi^l(r) P_l(\cos \theta) 
\label{el:At} \\
    j^t (r,\theta) &=& \frac{q}{4\pi}\frac{\delta(r-r_0)}{r_0^2}
\left(1-{2M\over R}\right)^{-1/2} \sum_{l=0}^{\infty} (2l+1) P_l(\cos \theta) \ .
\end{eqnarray}
Substituting these decompositions 
into Eq.\ (\ref{el:eq2}) we find that the latter becomes  
\begin{equation}
   \left \{ \begin{array}{ll}
   	\phi^l_{,rr}+\frac{2}{r}\phi^l_{,r}-\left(1-\frac{2M}{r}\right)^{-1}
\frac{l(l+1)}{r^2}\phi^l = 0 & \ \ \ r>R \\
	& \cr
	\phi^l_{,rr}+\frac{2}{r}\phi^l_{,r}-\frac{l(l+1)}{r^2}\phi^l = 
q\frac{(2l+1)}{r_0^2}\sqrt{1-\frac{2M}{R}}\delta(r-r_0) & \ \ \ r<R
\end{array} \right. \ \ .
\label{el:field}
\end{equation} 
The general solution for $r<R$ is a linear combinations of the basic solutions which
solve the corresponding homogeneous equation, i.e., a linear combination of $r^l$
and $r^{-l-1}$. The solution outside the shell is a linear combinations of 
$(r-2M)Q'_l[(r/M)-1]$ and $(r-2M)P'_l[(r/M)-1]$ 
\cite{israel,anderson,cohen-wald}. As in the case of scalar charge, 
we perform 
the coordinate transformation (\ref{eq:u}), re-express Eq.\ (\ref{el:eq2}) in terms
of 
$u$, integrate it from $u=R-\epsilon$ to $u=R+\epsilon$ and take the limit 
$\epsilon \rightarrow 0^+$. We end up with the same condition (\ref{bcR}) for 
$\phi^l_{,r}$ at $r=R$. The potential $A_t$ is solved in the same way as $\Phi$ 
in the scalar charge case. The matching conditions in this case are: (1) $\phi^l$ is 
continuous everywhere; (2) $\phi^l$ is finite at $r=0$ and vanishes at $r=\infty$; 
(3) $\phi^l_{,r}$ satisfies Eq.\ (\ref{bcR}) and
\begin{equation}
  \lim_{\epsilon \rightarrow 0^+} \left[ \phi^l_{,r}(r_0+\epsilon) -
\phi^l_{,r}(r_0-\epsilon) \right] = q\frac{2l+1}{r_0^2} \sqrt{1-{2M \over R}} \ ,
\label{el:match-r0}
\end{equation}
which comes from integrating Eq.\ (\ref{el:field}) across $r=r_0$.
The time component of the 4-vector potential $A_{\mu}$, the only non-vanishing 
component in our gauge, in a normalized basis is then
\begin{equation}
  A^{\hat{t}}(r,\theta) = \left \{ \begin{array}{lll} 
\sum_{l=0}^{\infty} G_l\, (r-2M)\, Q'_l\left( {r \over M}-1 \right)\,
P_l(\cos \theta) & \ \ \ r\geq R   & \ \ \ {\rm (region \ III)}   \\ & & \\ 
\sum_{l=0}^{\infty} q {r_0^l \over r^{l+1}} P_l(\cos \theta) 
+ \sum_{l=0}^{\infty} {q \over R} \left({r_0 \over R}
\right)^l \left({r \over R}\right)^l E^e_l \, P_l(\cos \theta) 
& \ \ \ r_0\leq r\leq R   & \ \ \ {\rm (region \ II)}   \\ & & \\    
\sum_{l=0}^{\infty} q {r^l \over r_0^{l+1}} P_l(\cos \theta) 
+ \sum_{l=0}^{\infty} {q \over R} \left({r_0 \over R}
\right)^l \left({r \over R}\right)^l E^e_l \, P_l(\cos \theta)
& \ \ \ r\leq r_0   & \ \ \ {\rm (region \ I)}             
\end{array}\right. \ \ ,
\end{equation}           
where 
\begin{equation}
  G_l = -q\frac{2l+1}{\sqrt{1-{2M \over r}}}\frac{M  r_0^l}{R^{l+3}}\left[
\left(1-{2M\over R}\right) Q_l''({R\over M}-1)-{M\over R} 
\left(l\sqrt{1-{2M\over R}}-1\right) Q_l'({R\over M}-1) \right]^{-1} \\
\end{equation}
and 
\begin{equation}
  E^e_l = \frac{{M\over R}\left[ (l+1)\sqrt{1-{2M\over R}}+1\right]Q_l'({R\over M}-1)+
\left(1-{2M\over R}\right) Q_l''({R\over M}-1)}{{M\over R}\left(l\sqrt{1-{2M\over R}}
-1\right) Q_l'({R\over M}-1)-\left(1-{2M\over R}\right) Q_l''({R\over M}-1)} \ .
\end{equation}                                        
In the neighborhood of the charge the potential then is given by
\begin{equation}   
A^{\hat{t}}(r,\theta) = 
\sum_{l=0}^{\infty} q \left[ {r^l \over r_0^{l+1}}
\Theta(r_0-r)+{r_0^l \over r^{l+1}}\Theta(r-r_0)\right] P_l(\cos \theta)
+ \sum_{l=0}^{\infty} {q \over R} \left({r_0 \over R}
\right)^l \left({r \over R}\right)^l E^e_l \, P_l(\cos \theta) \ \ .       
\end{equation}
The self force is computed by calculating the Lorentz force arising from this potential,
and evaluating it at the position of the charge. We thus find the bare force to be
\begin{equation}
f_r^{\rm bare}=
\sum_{l=0}^{\infty}\left[ -\frac{q^2}{2r_0^2}-\left({q\over R}\right)^2
 l\, E_l^e\, \left({r_0\over R}\right)^{2l-1}\right] \ \ .
\end{equation}                                           
This (bare) force diverges when the series is naively summed. 
\subsubsection{Regularization of the bare force}
Next, we regularize this
force using MSRP. Although MSRP has been developed only for scalar charges, it was shown in 
Refs. \cite{burko-cqg,burko-ajp} that it is also applicable for electric charges, at least in
simple cases. In particular, for an electric charge at rest in a locally-Minkowski
spacetime, the values of the regularization functions are known, and are given by 
$a_r=0=c_r$, $b_r=-q^2/(2r_0^2)$, and $d_r=0$. Consequently, the regularized, physical
self force is given by
\begin{equation}
f_r=
-\left({q\over R}\right)^2
\sum_{l=0}^{\infty} 
 l\, E_l^e\, \left({r_0\over R}\right)^{2l-1} \ \ .
\label{el:fmode}
\end{equation}        
\subsubsection{Properties of the regularized force}
When $E_l^e$ is expanded in $M/R$ we find 
\begin{equation}
   E_l^e =\frac{1}{2l+1}\left( {M\over R}\right)
+O\left({M\over
R}\right)^2 \ ,
\end{equation}
such that 
\begin{equation}
   f_r = -\left({q\over R}\right)^2\, {M\over R} \left[\sum_{l=1}^{\infty}
\frac{l}{2l+1} \left({r_0\over R}\right)^{2l-1} + O\left( {M\over R}\right)
\right] \ .
\end{equation}
To the leading order in $M/R$ this result can be written as 
\begin{equation}
f_r = -q^2\frac{G}{c^2}{M\over R^3} \frac{1}{2x_0}\left[
{1 \over 1-x_0^2}- {1\over 2x_0} \ln \left( {1+x_0 \over 1-x_0} \right)\right]  \ \ ,
\label{el:fin}
\end{equation}
where $x_0=r_0/R$. This force is directed toward the center, or away from the sphere, and
in this sense it is a repelling force. In Eq.\ (\ref{el:fin}) we re-introduced Newton's
constant $G$ and the speed of light $c$. This force is a 1st post-Newtonian effect.  

When the charge is close to the center, the self force is proportional to $r$, 
and the charge will oscillate with frequency 
\begin{eqnarray}
  \omega &=& \sqrt{\frac{1}{3}q^2\frac{G}{c^2}{M\over m R^4}} \\
	&=& 3.9\times 10^{-5} \frac{\rm rad}{\rm s} \left(\frac{q}{e}\right) \left( 
{M\over M_{\odot}} \right)^{1/2}  \left( {m \over m_e}\right)^{-1/2} 
\left({R\over 3\, {\rm km}}\right)^{-2} \ ,
\end{eqnarray}
where $m$ is the charge's mass, $e$ is electron's charge, $m_e$ is the electron's
mass, and $M_{\odot}$ is 
solar mass. This frequency is small, and is unlikely of being detected.  
Next we consider the force to the leading orders in $1-x_0$, and find that 
\begin{equation}
f_r = -q^2\frac{G}{c^2}{M\over R^3} \left \{ 
{1\over 4(1-x_0)}+{1\over 4}\ln(1-x_0)+{3\over 8}-{\ln 2\over 4}+
O[(1-x_0)\ln(1-x_0)]\right \} \ .
\label{el:lim-in}
\end{equation}
Again, this force diverges in the coincidence limit ($x_0\rightarrow 1^-$), but this
divergence only signifies the breakdown of our model at this limit. 

Fig.\ \ref{fig:elin} shows the self force $f_r$ as a 
function of $R/M$ computed by full expression (\ref{el:fmode}) (solid line) 
and the small $M/R$ expression (\ref{el:fin}) (dashed line). The result is 
very similar to the scalar charge case (cf.\ Fig.\ \ref{fig:scin}).
It can be shown that 
\begin{equation}
  f_r \rightarrow -\frac{q^2}{R^2}\frac{G}{c^2} \frac{2x_0-x_0^3}{(1-x_0^2)^2}
\label{el:asyin}
\end{equation}
as  $R/M \rightarrow 2$.
\subsubsection{Alternative regularization procedure}
Similarly to the direct application of the Comparison Axiom in the scalar case, we can apply
the electric-field Comparison Axiom \cite{quinn-wald-97} here analogously. The electric
field Comparison Axiom is nearly identical to the scalar-field Comparison Axiom. In fact,
all that we need to change in its definition is to replace the scalar fields $\phi$ and
${\tilde \phi}$ with the Maxwell fields $F_{\mu\nu}$ and ${\tilde F}_{\mu\nu}$,
respectively, and replace Eq.\ 
(\ref{sc:ax}) with 
\begin{equation}
f_{\mu}-{\tilde f}_{\mu}=\lim_{x\to 0}q\left<F_{\mu\nu}-{\tilde
F}_{\mu\nu}\right>_{x}u^{\nu} . \
\end{equation}
In order to apply the Comparison Axiom we write the potential in the form 
\begin{equation}
  A^{\hat{t}}(r,\theta) = \left \{ \begin{array}{ll}
        \sum_{l=0}^{\infty} G_l\, (r-2M)\, Q'_l\left( {r \over M}-1 \right)\,
P_l(\cos \theta) & \ \ \ r\geq R \\
        & \\
        \Phi_c(r,\theta) + \Delta \Phi_e(r,\theta) & \ \ \ r \leq R
\end{array}\right. \ \ ,
\end{equation}
where $\Phi_c$ is the Coulomb field (\ref{coulumb}), which is the potential which generates 
${\tilde F}_{\mu\nu}$,  and 
\begin{equation} 
\Delta \Phi_e(r,\theta) = \sum_{l=0}^{\infty} {q \over R} \left({r_0 \over R}
\right)^l \left({r \over R}\right)^l E^e_l \, P_l(\cos \theta) \ \ .             
\end{equation}
is the correction term due to the finite size of the Minkowski patch of spacetime. 
Applying the Comparison Axiom, we find that the self force is given by 
\begin{equation}
   f_r = -\left({q\over R}\right)^2\sum_{l=1}^{\infty} l\, E_l^e\, 
\left({r_0\over R}\right)^{2l-1} \ ,
\end{equation}
which is identical to Eq.\ (\ref{el:fmode}).

\subsection{Electric charge outside the shell}\label{sec:electric-out}
\subsubsection{Derivation of the bare force}
Next, we shall study the case where the electric charge $q$ is outside the shell, i.e., the
case where $r_0>R$. The time-component of the 4-current density is now decomposed as
\begin{equation}
  j^t=\frac{q}{4\pi}\, \frac{\delta(r-r_0)}{r_0^2}\sum_{l=0}^{\infty} (2l+1) P_l(\cos
\theta)
\ \ .
\end{equation}
With $A_t$ decomposed as in (\ref{el:At}), the Maxwell equation (\ref{el:eq}) 
becomes
\begin{equation}
   \left \{ \begin{array}{ll}
	\phi^l_{,rr}+{2\over r}\phi^l_{,r}-\left(1-{2M\over r}\right)^{-1} 
\frac{l(l+1)}{r^2}\phi^l=q(2l+1)\frac{\delta(r-r_0)}{r^2} & \ \ \ \ \ r>R \\
& \\
	\phi^l_{,rr}+{2\over r}\phi^l_{,r}-\frac{l(l+1)}{r^2}\phi^l=0 & 
\ \ \ \ \ r<R \end{array} \right. \ \ .
\end{equation}
Hence, $\phi^l$ can be written as
\begin{equation}
   \phi^l(r) = \left \{ \begin{array}{lll}
  A_l (r-2M) Q'_l\left( {r\over M}-1\right) & \ \ \ \ \ r\geq r_0 & \ \ \ \ {\rm (region \
III)} 
\\ & \\
  C_l (r-2M) P'_l\left( {r\over M}-1\right)+D_l(r-2M) Q'_l\left( {r\over M}-1\right)
& \ \ \ \ \ R \leq r \leq r_0 & \ \ \ \ {\rm (region \ II)}                                                                
\\ & \\
  B_l r^l & \ \ \ \ \ r \leq R & \ \ \ \ {\rm (region \ I)}
\end{array} \right. 
\end{equation}
when $l \neq 0$. When $l=0$ the solution has the form 
\begin{equation}
  \phi^0(r)= \left \{ \begin{array}{lll} 
	{A_0 \over r} & \ \ \ \ \ r \geq r_0 & \ \ \ \ {\rm (region \ III)}                                                               
\\ & \\
	C_0+{D_0 \over r} & \ \ \ \ \ R \leq r \leq r_0 & \ \ \ \ {\rm (region \ II)}  
\\ & \\
	B_0 & \ \ \ \ \ r \leq R & \ \ \ \ {\rm (region \ I)} 
\end{array} \right. \ \ \ \ \ .
\end{equation}
$A_l$, $B_l$, $C_l$ and $D_l$ are constants to be determined by the 
matching conditions, which are the same as the case when the electric charge is 
inside the shell, except that eq.\ (\ref{el:match-r0}) is replaced by 
\begin{equation}
  \lim_{\epsilon \rightarrow 0^+} \left[ \phi^l_{,r}(r_0+\epsilon) -
\phi^l_{,r}(r_0-\epsilon) \right] = q\frac{2l+1}{r_0^2} \ .
\end{equation}

Straightforward calculations yield 
\begin{equation}
  A_t(r,\theta)= \left \{ \begin{array}{ll}
       - {q \over r} +
\sum_{l=1}^{\infty} {q\over M^3}\frac{2l+1}{l(l+1)}(r-2M)(r_0-2M) P_l(\cos \theta)  
\left[P_l'(z_0) Q_l'(z)+\tilde{E}^e_lQ'_l(z_0)Q'_l(z)\right]
& \ \ \ \ \ r \geq r_0 
\\ & \\
   - {q \over r_0} +
\sum_{l=1}^{\infty} {q\over M^3}\frac{2l+1}{l(l+1)}(r-2M)(r_0-2M) P_l(\cos \theta)
\left[P_l'(z) Q_l'(z_0)+\tilde{E}^e_lQ'_l(z_0)Q'_l(z)\right]                     
& \ \ \ \ \ R \leq r \leq r_0 
\\ & \\
 - {q \over r_0} + 
\sum_{l=1}^{\infty} 
q\frac{M r_0}{R^{l+3}}
\frac{(2l+1)\left(1-{2M\over r_0}\right) Q'_l(z_0)}{\left(1-{2M\over
R}\right) Q''_l(Z)-{M\over R}\left(l \sqrt{1-{2M\over R}}-1\right)
Q'_l(Z)}\ r^l \ P_l ( \cos \theta )       
& \ \ \ \ \ r \leq R 
\end{array} \right. \ \ \ \ \ .
\end{equation}                                                
Here, 
\begin{equation}
 \tilde{E}^e_l=\frac{-(1-{2M\over R})P''_l\left({R\over M}-1\right)+{M\over R}
\left(l \sqrt{1-{2M\over R}}-1\right)P'_l\left({R\over M}-1\right)}
{(1-{2M\over R})Q''_l\left({R\over M}-1\right)-{M\over R}
\left(l \sqrt{1-{2M\over R}}-1\right)Q'_l\left({R\over M}-1\right)} \ ,
\end{equation}                 
and we have used the Wronskian $W[P_n'(z),Q_n'(z)]=n(n+1)/(z^2-1)^2$ 
to simplify the expressions.
In the neighborhood of the charge the potential can thus be written as 
\begin{eqnarray}
   A_t &=& - {q\over r}\Theta(r-r_0)-{q\over r_0}\Theta(r_0-r)+
\sum_{l=1}^{\infty} {q\over M^3}\frac{2l+1}{l(l+1)}(r-2M)(r_0-2M) P_l(\cos \theta)
\times \cr
& & \left[ P_l'(z) Q_l'(z_0)\Theta(r_0-r)
+P_l'(z_0) Q_l'(z)\Theta(r-r_0)+\tilde{E}^e_lQ'_l(z_0)Q'_l(z)\right] \ .
\end{eqnarray}                    
The (bare) force is given formally by
\begin{eqnarray}
  f_r^{\rm bare} &=& qF_{r \mu} u^{\mu} \cr
   &=& q \left(1-{2M\over r_0}\right)^{-1/2} A_{t,r} 
\ ,
\end{eqnarray}                                     
which of course diverges. 
\subsubsection{Regularization of the bare force}
The application of MSRP to this case is not straightforward,
because the analytical value of $b_r$ is as yet unknown. However, it was shown in Ref.\ 
\cite{burko-cqg} that it can be computed numerically, by studying the large-$l$ behavior of
the
modes of the bare force. It was also found numerically in  Ref.\ \cite{burko-cqg} that for an
electric charge at rest in Schwarzschild spacetime the values of the other
regularization
functions are given by $a_r=0=c_r$ and that $d_r=0$.  When MSRP is applied, the
regularized, physical self force in an orthonormal frame is found to be given
by 
\begin{eqnarray} 
f_{\hat{r}} &=& 
q^2\frac{M}{r_0^3} \cr
&+&\frac{q^2r_0}{M^3}\left(1-{2M\over r_0}\right)\sum_{l=1}^{\infty}
\frac{2l+1}{l(l+1)}\tilde{E}^e_l \left[ Q'_l\left({r_0\over M}-1\right)\right]^2 \cr
 &+&\frac{q^2 r_0^2}{M^4}\left(1-{2M\over r_0}\right)^2\sum_{l=1}^{\infty}
\frac{2l+1}{l(l+1)}\tilde{E}^e_l Q_l'\left({r_0\over M}-1\right)Q_l''\left(
{r_0\over M}-1\right) \ .
\label{el:fout}
\end{eqnarray}                
\subsubsection{properties of the regularized force}
To leading order in $M/R$, the  self force is reduced to the
following simple expression 
\begin{equation}
  f_{\hat{r}} = \frac{q^2 M}{r_0^3} \sum_{l=0}^{\infty} \frac{l+1}{2l+1}\left(
{R\over r_0}\right)^{2l} \ , 
\end{equation}
which can be written as 
\begin{equation}
f_{\hat{r}}  = q^2 \frac{G}{c^2} \frac{M}{2r_0^3}\left[ \frac{1}{1-y_0^2}+\frac{1}{2y_0}\ln
\left( \frac{1+y_0}{1-y_0}\right) \right] \ \ ,
\label{el:finout}
\end{equation}
where $y_0=R/r_0$. In the last equation we re-introduced Newton's constant $G$ and the
speed of light $c$. The effect of the finite size of the shell is to increase the
self force repulsion which was found first by Smith and Will for the case where the
source of the gravitational field is a (Schwarzschild) black hole 
\cite{smith-will-80}. To leading orders in $1-y_0$ we find that the force is given by 
\begin{equation}
  f_{\hat{r}}=q^2 \frac{G}{c^2} \frac{M}{r_0^3}\left \{ {1\over 4(1-y_0)}-{1\over 4}
\ln(1-y_0)+{1\over 8}+{\ln 2\over 4}+O[(1-y_0)\ln(1-y_0)]\right \} \ .
\end{equation}
This force diverges in the coincidence limit ($y_0 \rightarrow 1^-$), again, due to
the breakdown of the mathematical model. 

Fig.\ \ref{fig:elout} shows (a) the total self force and (b) the correction of 
the self force due to the finite Minkowski patch [i.e.\ $\Delta f_{\hat{r}}$ in 
Eq.\ (\ref{el:fdiff})] 
in unit of $q^2G/(c^2 R^2)$ as a 
function of $R/M$ based on the full expression (\ref{el:fout}) and the small 
$M/R$ formula (\ref{el:finout}). The charge is placed at $r_0/R=2$. We can see 
that the situation is very similar to the scalar case 
(c.f.\ Fig.\ {\ref{fig:scout}). 
The self force reduces to the Will-Smith result in 
the limit $R/M\rightarrow 2$, as expected.

Finally, Fig.\ \ref{fig:x0}b shows the magnitude of the self force in unit of
$q^2G/(c^2 R^2)$ as a function of $r_0/R$ for $R=2.5M$. We can see that 
the small $M/R$ expressions (dashed line) deviates significantly from 
the full expression (solid line) when 
the charge is inside the shell, but is quite accurate when it is outside 
the shell. The reason is that the self force is dominated by the 
Smith-Will force, which are present in both the full and the small $M/R$ 
expressions, when the charge is outside the shell.
The self force decreases as the charge moves away from the shell, but 
diverges at $r_0=R$ (see the above discussion).
\subsubsection{Alternative regularization procedure}                     
Again, we can apply the Comparison Axiom directly. We thus write the potential as 
\begin{equation}
   A_t = \left \{ \begin{array}{ll} 
  	A_t^{\rm sch}+\Delta A_t & \ \ \ \ \ r\geq R \\ & \\
	B_l r^l P_l(\cos \theta) &  \ \ \ \ \ r\leq R
\end{array} \right.
\end{equation}
where 
\begin{eqnarray}
   A_t^{\rm sch}&=&-{q\over r}\Theta(r-r_0)-{q\over r_0}\Theta(r_0-r)+ 
\sum_{l=1}^{\infty} {q\over M^3}\frac{2l+1}{l(l+1)}(r-2M)(r_0-2M) P_l(\cos \theta)
\times \cr
& & \left[ P_l'\left({r\over M}-1\right) Q_l'\left({r_0\over M}-1\right)\Theta(r_0-r) 
+P_l'\left({r_0\over M}-1\right) Q_l'\left({r\over M}-1\right)\Theta(r-r_0) \right]
\end{eqnarray}
is the potential if the shell is replaced by a Schwarzschild black hole of the same 
mass \cite{burko-cqg} (i.e., a point-like source for the gravitational field), which is the
origin for ${\tilde F}_{\mu\nu}$, and
\begin{equation}
  \Delta A_t={q\over M^3} (r_0-2M)(r-2M)\sum_{l=1}^{\infty} 
\frac{2l+1}{l(l+1)} \tilde{E}^e_l Q_l'\left({r_0\over M}-1\right) 
Q_l'\left({r\over M}-1\right) P_l(\cos \theta) \ \ 
\end{equation}
is the correction for the potential because of the finite size of the Minkowski patch
of spacetime. Here,
\begin{equation}
  B_l= \left \{ \begin{array}{ll}
	q(2l+1)M\frac{r_0}{R^{l+3}}\left(1-{2M\over r_0}\right)
\frac{Q'_l\left({r_0\over M}-1\right)}{\left(1-{2M\over R}\right) Q''_l
\left({R\over M}-1\right)-{M\over R}\left(l \sqrt{1-{2M\over R}}-1\right)
Q'_l\left({R\over M}-1\right)} & \ \ \ \ \ l\neq 0 \\ & \\
  	-{q\over r_0} & \ \ \ \ \ l=0 \end{array} \right. \ \ .
\end{equation}
The difference in the self forces between the actual and the ``tilde'' spacetimes is
then given by  
\begin{eqnarray}
  f_r - {\tilde f}_r&=& q(F_{r \mu}-{\tilde F}_{r \mu}) u^{\mu} \cr
   &=& q \left(1-{2M\over r_0}\right)^{-1/2} (\Delta A_{t})_{,r} 
\ \ . 
\end{eqnarray}
This difference then equals 
\begin{eqnarray}
f_r - {\tilde f}_r\equiv \Delta f_r
&=& q \left(1-{2M\over r_0}\right)^{-1/2}\Delta A_{t,r} \cr
f_{\hat{r}}
  &=& \frac{q^2r_0}{M^3}\left(1-{2M\over r_0}\right)\sum_{l=1}^{\infty} 
\frac{2l+1}{l(l+1)}\tilde{E}^e_l \left[ Q'_l\left({r_0\over M}-1\right)\right]^2 \cr
 & & +\frac{q^2 r_0^2}{M^4}\left(1-{2M\over r_0}\right)^2\sum_{l=1}^{\infty}
\frac{2l+1}{l(l+1)}\tilde{E}^e_l Q_l'\left({r_0\over M}-1\right)Q_l''\left(
{r_0\over M}-1\right) \ .
\label{el:fdiff}
\end{eqnarray}
Recall now that $f_{\hat{r}}=\Delta f_{\hat{r}} +{\tilde f}_{\hat{r}}$, 
where ${\tilde f}_{\hat{r}}$ is the
Smith-Will force given by ${\tilde f}_{\hat{r}}=q^2(M/r_0^3)$. 
When combined, we recover the self force $f_{\hat{r}}$, which is identical to 
Eq.\ (\ref{el:fout}). 

\section{Heuristic viewpoint on the origin of the self force}\label{heuristic}

The self force experienced by a static electric charge in the presence of the 
shell can be interpreted as a result of the interaction between the charge's 
electric field and the shell's gravitational field. In this section, we 
shall consider the self force in the small $M/R$ limit, i.e. Eqs.\ (\ref{el:fin}) 
and (\ref{el:finout}), by a heuristic argument. It is the linearization in 
$M/R$ which allows us to obtain the solution very simply. 

Consider a static electric charge $q$ at the origin of the coordinate system. 
Spacetime is then described by the Reissner-Nordstr\"{o}m metric
\begin{equation}
  ds^2=-\left(1-{2m\over r}+{q^2\over r^2}\right)\,dt^2+\left(1-{2m\over r}+
{q^2\over r^2}\right)^{-1}\,dr^2+
r^2 (\,d\vartheta^2+\sin^2 \vartheta \,d\varphi^2) \ \ ,
\label{re-norm:metric}
\end{equation}
where $m$ is the mass of the charge. A particle of mass $\mu$ fixed at $r$ will 
experience a ``gravitational force'' $f^{\alpha}=-\mu \, D u^{\alpha}/D\tau$, 
where $u^{\alpha}$ is the 4-velocity and $D/D\tau$ denotes covariant derivative 
[compatible with the metric (\ref{re-norm:metric})]
with respect to the particle's proper time. Due to the symmetry of the setup, 
the only non-vanishing
component  of $f^{\alpha}$, in an orthonormal basis, is 
\begin{eqnarray}
   f^{\hat{r}} &=& -\mu (u^t)^2\, \Gamma^r_{tt} \left(1-{2m\over r}+{q^2\over r^2} 
\right)^{-1/2} \cr
  &=& -{\mu m\over r^2}+{\mu q^2\over r^3}
\label{grav:force}
\end{eqnarray}
to leading order in $m$ and $q^2$, 
where $\Gamma^{\alpha}_{\beta \gamma}$ are the connection coefficients. The first 
term is the usual (attractive) gravitational force, whereas the second term, which
is a repulsive  force, comes from the stress-energy tensor of the electric field. 
In what follow we shall ignore the former, as we are interested only in the 
electric-field interaction and not in the direct gravitational force. 
Because of the staticity of the problem, the two-body system (charge and 
massive particle) conserves linear momentum. Consequently, Newton's third law is 
applicable for this two-body system. (In general, when radiation is present, Newton's 
third law is inapplicable.)  
It follows from Newton's third law that the static electric charge also experiences
this repulsive force  $q^2 \mu/r^3$ apart from the usual attractive gravitational
force. This additional repulsive force on the static charge is then interpreted as
a result of the  interaction between the charge's electric field and the point mass
$\mu$, i.e., it is the self force. Notice that this result is the same as the self
force acting on a static electric charge in Schwarzschild spacetime computed
by Smith and
Will \cite{smith-will-80} (notice, however, that by this argument we only find the
leading order term in $\mu$. The Smith-Will force, however, is an exact result).  

Now, suppose the charge is surrounded by a spherical shell of radius $R$ and 
mass $M$ (Fig.\ \ref{fig:geo}a). In the small $M/R$ and test charge limits, the 
``gravitational force'' acting on the shell by the charge's electric field is 
equal to the sum of the forces acting on each particle on the shell. We express 
the mass $\mu$ of a small element of the shell by $\mu=M\,d^2x/(4\pi R^2)$, where 
$\,d^2x$ is an area element of the shell. From symmetry, 
it is clear that the total force is aligned on the 
$z$-axis. We thus project all the 
contributions to the total force on this axis, 
such that the vector summation becomes 
trivial. The resultant force is along 
the $z$-axis and is given by
\begin{eqnarray}
  F^{\hat{z}} &=& \frac{M}{4\pi R^2}\int_{shell} \,dx^2\, \frac{q^2}{r^3}
\cos(\theta+\xi) \cr
  	&=& \frac{q^2 M}{2}\int_{-1}^{1} \frac{R\cos \theta -r_0}
{(r_0^2+R^2-2r_0 R\cos \theta)^2} \,d (\cos \theta) \cr
  &=& q^2 {G\over c^2} {M\over R^3} {1 \over 2x_0}\left[ 
\frac{1}{1-x_0^2}-\frac{1}{2x_0}\ln\left( \frac{1+x_0}{1-x_0}\right)\right] \ ,
\label{force:heui}
\end{eqnarray}
where the angles $\theta$ and $\xi$ are defined in Fig.\ \ref{fig:geo}a, and
\begin{equation}
  r=\sqrt{r_0^2+R^2-2r_0 R\cos \theta} \ \ \ \mbox{and} \ \ \ 
\cos (\theta+\xi) = \frac{R\cos \theta -r_0}{r} \ \ . 
\end{equation}
We re-introduced Newton's constant $G$ and the speed of light $c$ in the last 
expression of (\ref{force:heui}). The self force experienced by the charge 
is of the same magnitude but directs towards the center of the shell, which is 
exactly the same as Eq.\ (\ref{el:fin}). 

When the charge is outside the shell, we have (Fig.\ \ref{fig:geo}b)
\begin{eqnarray}
  F^{\hat z} &=& -\frac{M}{4\pi R^2} \int_{shell} \,dx^2 \, {q^2\over r^3}
\cos \chi \cr
      &=& -\frac{q^2 M}{2r_0^3} \int_{-1}^1 \frac{1-y_0 \cos \theta}
{(1+y_0^2-2y_0\cos \theta)^2} \,d(\cos \theta) \cr
      &=& -q^2 \frac{G}{c^2} \frac{M}{2r_0^3}\left[ \frac{1}{1-y_0^2}+\frac{1}{2y_0}
\ln \left( \frac{1+y_0}{1-y_0}\right) \right] \ ,
\label{force:heuo}
\end{eqnarray}
where we have used the expressions 
\begin{equation}
  r=\sqrt{R^2+r_0^2-2r_0 R \cos \theta} \ \ \ \mbox{and} \ \ \ 
\cos \chi = \frac{r_0-R\cos \theta}{r} \ ,
\end{equation}
and re-introduced $G$ and $c$ in the last expression of (\ref{force:heuo}). Hence,
the self force is $-F^{\hat z}$ (repel the shell), exactly the same as 
in Eq.\ (\ref{el:finout}).

We can carry out the above calculation also to the scalar charge case. 
Consider a scalar test charge at the origin of the coordinates. Recall that 
we are interested here in the self force, and not in the usual gravitational 
attraction due to the particle's mass. Hence, in what follows we ignore the mass
of the scalar charge. Next, we write the coupled Einstein-Klein-Gordon equations, 
\begin{eqnarray}
 G_{\mu \nu} &=& 8\pi T_{\mu \nu} \cr
 \Box \Phi &=& -4\pi \rho \ ,
\label{ekg}
\end{eqnarray}
where $\Box$ denotes the covariant wave operator and $\rho$ is the charge density
given by Eq.\ (\ref{charge-density}) for a static charge at the origin, 
and look for a static, spherically symmetric solution. Here, $T_{\mu \nu}$ 
is the stress-energy tensor of a massless scalar field, which is  given by              
\begin{equation}
   T_{\mu \nu}={1\over 4\pi} \left( \Phi_{,\mu}\Phi_{,\nu}-{1\over 2} g_{\mu \nu}
g^{\alpha \beta} \Phi_{,\alpha}\Phi_{,\beta}\right) \ .
\label{Tab:sc}
\end{equation}                  
The Einstein equations then reduce to 
\begin{eqnarray}
 R_{\mu \nu} &=& 2\Phi_{,\mu}\Phi_{,\nu} \cr
 \Box \Phi &=& -4\pi \rho \ ,
\label{ekg1}
\end{eqnarray}           
whose solution is given by 
\begin{equation}
\,ds^2=-\,dt^2+\frac{\,dr^2}{1+\frac{q^2}{r^2}}+r^2 \left(\,d\vartheta^2 +
\sin^2 \vartheta \,d\varphi^2\right) 
\label{metric:sc}
\end{equation}              
and 
\begin{equation}
\Phi_{,r}=-\frac{q}{r^2}\frac{1}{\sqrt{1+\frac{q^2}{r^2}}} \ .
\label{sc:sc2}
\end{equation}
%
%
Hence the ``gravitational force'' experienced by a static point mass $\mu$ is 
$f^{\alpha}=-\mu \,D u^{\alpha}/\,D\tau=-\mu (u^t)^2\,
\Gamma^{\alpha}_{tt}=0$, since 
$\Gamma^{\alpha}_{tt}=0$ for the metric (\ref{metric:sc}). $D/\,D\tau$ here
denotes covariant differentiation compatible with the metric
(\ref{metric:sc}). 
Thus we conclude that the self force is zero to leading order in $\mu$. 
Consequently, after integration over the shell, the self force is zero also to
linear 
order in $M$. This result is in accord with our 
previous calculation that the self force in the scalar charge case is a second 
post-Newtonian effect.

We can also make the following arguments for the direction of the self force. For
concreteness, consider the case of an electric charge inside the shell. The $4\pi$ 
electric field lines  near the charge have the usual distribution in space. However,
outside the shell they are distorted due to the curvature of space. Specifically,
the electric field lines are closer than what they would be if spacetime were flat.
The reason is that the ratio of the circumference and the (proper) radius is smaller
than $2\pi$. When the charge is off the center of the sphere, the curvature effect
is stronger outside the shell at the side closer to the charge. Because of the
stress in the electrostatic field (one has to put in energy to squeeze electric
field lines), this results in a force acting on the charge in the direction of the
center of the sphere. An alternative viewpoint is the following. The electrostatic
field is accompanied by a stress-energy tensor which gives rise to an effective
metric which is similar to the Reissner-Nordstr\"{o}m metric without the mass term.
This metric induces repulsive (or anti-) gravity, which is the origin for the
repulsive force acting on the charge. (The usual Reissner-Nordstr\"{o}m geometry is
attractive at large distances because of the mass term. It is, however, repulsive at
short distances, where the mass term is small compared with the charge term in the
metric. This happens, nevertheless, only deep inside the Reissner-Nordstr\"{o}m
black hole, and is responsible for the phenomenon of gravitational bounce 
\cite{novikov-66}. The occurrence of gravitational bounce in actuality is uncertain
because of the inner-horizon instability of realistic black holes 
\cite{penrose-68}. For recent reviews see \cite{burko-ori,brady}.) 
Consequently, the self force on the charge directs toward the
center of the shell.

\begin{figure}
\epsfig{file=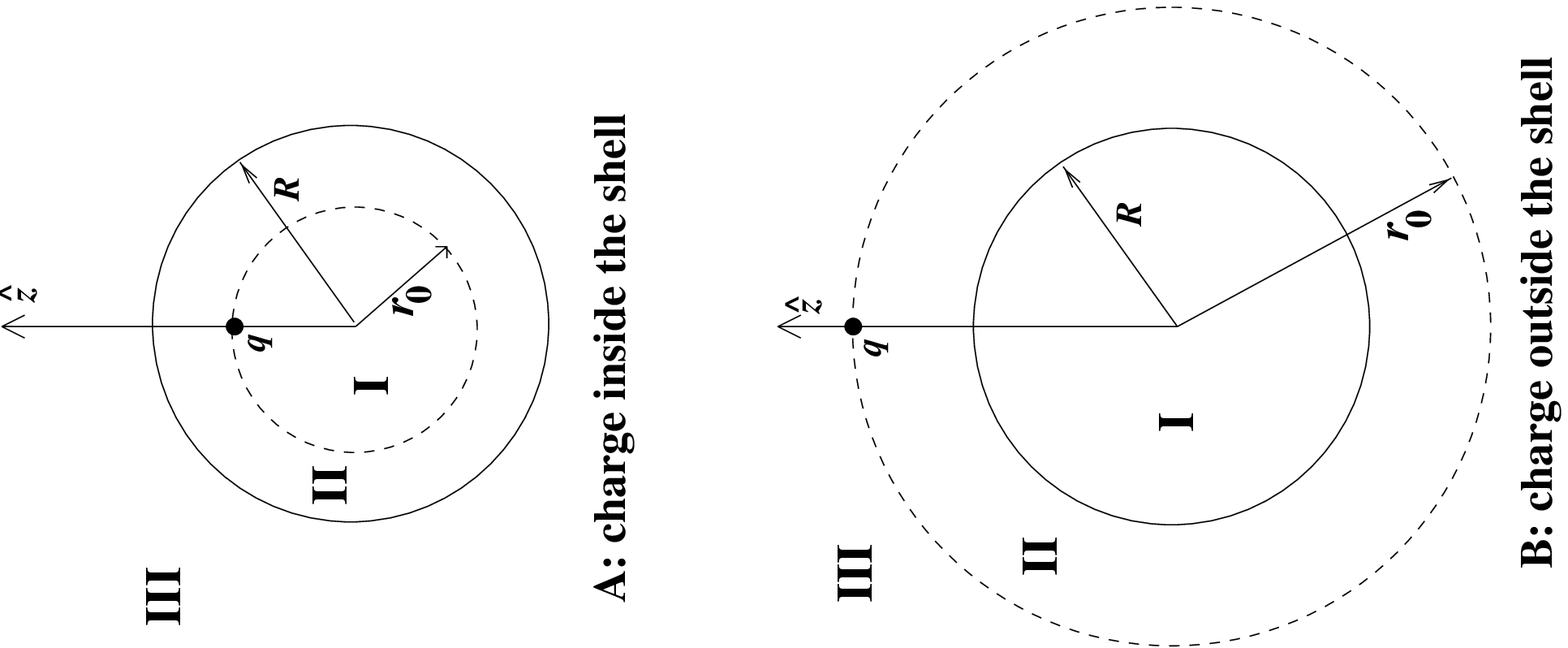,width=10cm,angle=270}
\caption{A test charge (scalar or electric) $q$ is placed (a) inside and 
(b) outside a spherical shell of mass $M$ and radius $R$. In both cases the particle is at
$r=r_0$, and without loss of generality the particle is positioned on the $\hat z$-axis. See
the text for more details.} 
\label{fig1}
\end{figure}

\begin{figure}
\epsfig{file=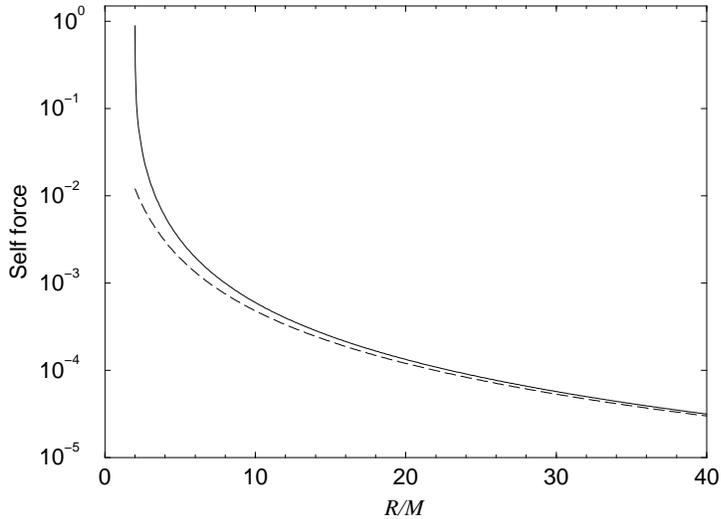,width=7cm,angle=270}
\caption{The magnitude of the self force $|f_{\hat{r}}|$, in units of $q^2G^2/(R^2 c^4)$, 
experienced by a static, test, scalar charge as a function of $R/M$ computed by 
the full expression Eq.\ 
(\ref{sc:sf}) (solid line) and the small $M/R$ expansion (\ref{sc:lin}) (dashed 
line). 
The charge is placed inside the shell at $r_0/R=1/2$, and the self force approaches 
the limit $(8/9)q^2G^2/(R^2 c^4)$ predicted by (\ref{sc:asyin}) 
when $R/M \rightarrow 2$.}
\label{fig:scin}
\end{figure}

\begin{figure}
\epsfig{file=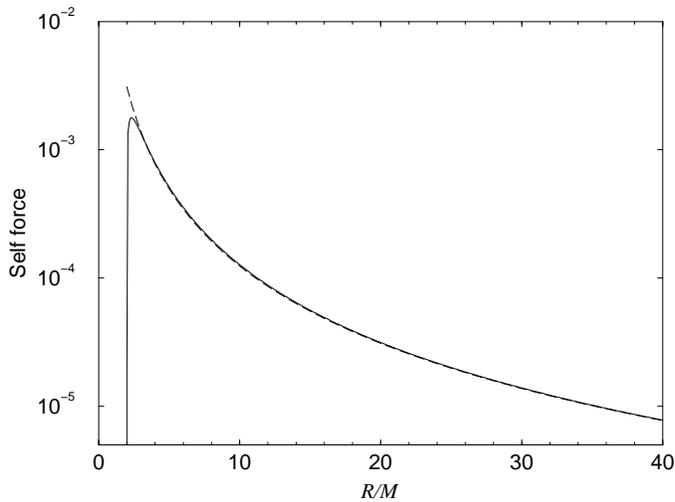,width=7cm,angle=270}
\caption{Same as Fig.\ \ref{fig:scin} but the scalar charge is placed outside 
the shell at $r_0/R=2$. Solid line is computed by the full expression Eq.\
(\ref{sc:sf2}); dashed line is computed
by the small $M/R$ formula (\ref{scout:lin}). The self force vanishes in the 
limit $R/M \rightarrow 2$.}
\label{fig:scout}
\end{figure}

\begin{figure}
\epsfig{file=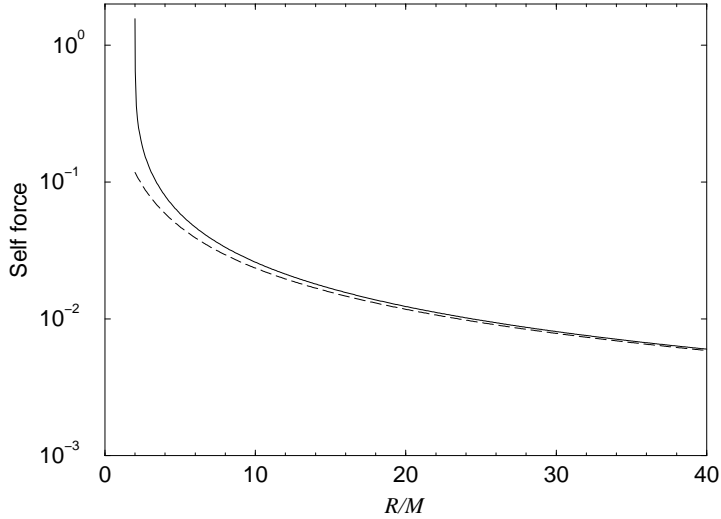,width=7cm,angle=270}
\caption{The magnitude of the self force $|f_{\hat{r}}|$, in units of
$q^2G/(R^2 c^2)$,
experienced by an electric charge as a function of $R/M$, computed by the full expression Eq.\ 
(\ref{el:fmode}) (solid line) and the small $M/R$ formula (\ref{el:fin}) 
(dashed line).
The charge is placed inside the shell at $r_0/R=1/2$, and the self force 
approaches the limit $(14/9)q^2G/(R^2 c^2)$ predicted by (\ref{el:asyin}) when 
$R/M \rightarrow 2$.}
\label{fig:elin}
\end{figure}

\begin{figure}
\epsfig{file=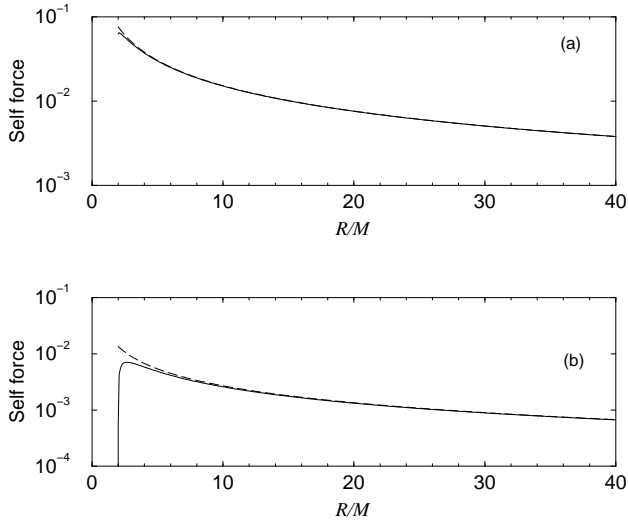,width=7cm,angle=270}
\caption{Same as Fig.\ \ref{fig:elin} but the electric charge is placed outside
the shell at $r_0/R=2$. The upper panel (a) shows the total self force. The lower 
panel (b) shows the correction to the self force due to the finite size of
the 
Minkowski patch [i.e.\ $\Delta f_{\hat{r}}$ in Eq.\ (\ref{el:fdiff})].
Solid lines are calculated by the full expression Eq.\  
(\ref{el:fout}); dashed lines are calculated based 
on the small $M/R$ formula (\ref{el:finout}). 
The self force approaches the Smith-Will result when $R/M \rightarrow 2$.}
\label{fig:elout}
\end{figure}

\begin{figure}
\epsfig{file=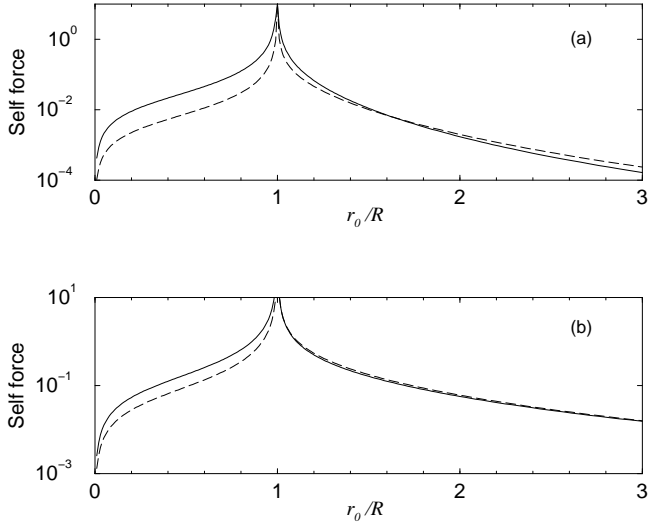,width=7cm,angle=270}
\caption{The magnitude of the self force $|f_{\hat r}|$ experienced by (a) a scalar 
charge [in unit of $q^2G^2/(c^4 R^2)$] and (b) an electric charge [in unit 
of $q^2G/(c^2R^2)$] as a function of the charge's position $r_0$. The 
radius of the shell is set to $R=2.5M$.  Solid 
lines are calculated by corresponding full expressions and dashed lines are calculated by 
the appropriate expressions for small $M/R$.}
\label{fig:x0}
\end{figure}

\begin{figure}
\epsfig{file=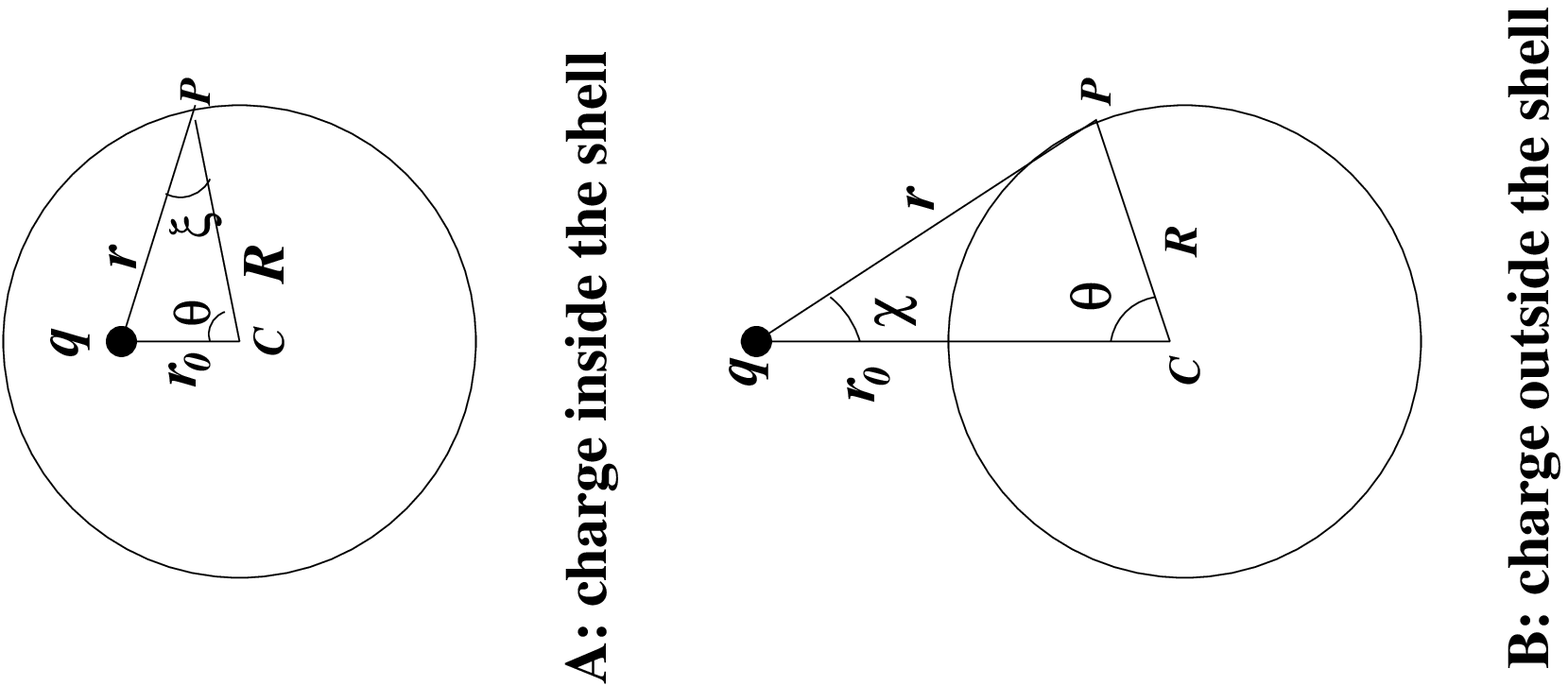,width=10cm,angle=270}
\caption{Geometry of the charge and the shell in Sec.\ \ref{heuristic}: 
(a) the charge is inside the shell;
(b) the charge is outside the shell.
$R$ is the shell's radius; $r_0$ is the distance between the charge $q$ and the 
shell's center $C$. $P$ is an arbitrary point on the shell. $r$ is the distance 
between the charge $q$ and point $P$. $\theta$ is the angle between lines $Cq$ and 
$CP$; $\xi$ (in the upper figure) is the angle between lines $qP$ and $CP$, and 
$\chi$ (in the lower figure) is the angle between lines $qC$ and $qP$.}
\label{fig:geo}
\end{figure}

\section*{acknowledgments}
We thank Jeremy Heyl, Amos Ori, and Kip Thorne for invaluable discussions. L.M.B. \ 
wishes to thank the Technion Institute of Theoretical Physics,
where part of this research was done, for hospitality. At Caltech this
research was supported by NSF grants AST-9731698 and PHY-9900776 and by NASA grant
NAG5-6840.

\end{document}